\RequirePackage{fix-cm}
\documentclass[smallextended]{svjour3}       
\smartqed  
\usepackage{algorithm2e}
\usepackage{algpseudocode}
\usepackage{amsmath}
\usepackage{amssymb}
\usepackage{color}
\usepackage{graphicx}
\usepackage{gnuplot-lua-tikz}
\usepackage{hyperref}

\newcommand{\Aff}[2]{#1}

\newcommand{\MyEmph}[1]{\textbf{#1}}

\newcommand{\In}{\ \ \ }   

\newcommand{\R}{\ensuremath{\mathbb{R}}}

\newcommand{\dz}{\, \mathrm{d}z}

\DontPrintSemicolon		

\begin{document}

\title{Benefits from using mixed precision computations in the
       ELPA-AEO and ESSEX-II eigensolver projects%
  \thanks{This work has been supported by the
          Deutsche Forschungsgemeinschaft through the
          priority programme~$1648$ ``Software for Exascale Computing''
          (SPPEXA) under the project ESSEX-II
          and by the Federal Ministry of Education and Research
          through the project ``Eigenvalue soLvers for Petaflop
          Applications -- Algorithmic Extensions and Optimizations''
          (ELPA-AEO) under Grant No.~01H15001.}}

\titlerunning{Mixed precision in the ELPA-AEO and ESSEX-II projects}

\author{%
  Andreas Alvermann \and
  Achim Basermann \and
  Hans-Joachim Bungartz \and
  Christian Carbogno \and
  Dominik Ernst \and
  Holger Fehske \and
  Yasunori Futamura \and
  Martin Galgon \and
  Georg Hager \and
  Sarah Huber \and
  Thomas Huckle \and
  Akihiro Ida \and
  Akira Imakura \and
  Masatoshi Kawai \and
  Simone K\"ocher \and
  Moritz Kreutzer \and
  Pavel Kus \and
  Bruno Lang \and
  Hermann Lederer \and
  Valeriy Manin \and
  Andreas Marek \and
  Kengo Nakajima \and
  Lydia Nemec \and
  Karsten Reuter \and
  Michael Rippl \and
  Melven R\"ohrig-Z\"ollner \and
  Tetsuya Sakurai \and
  Matthias Scheffler \and
  Christoph Scheurer \and
  Faisal Shahzad \and
  Danilo Simoes Brambila \and
  Jonas Thies \and
  Gerhard Wellein
}

\authorrunning{Andreas Alvermann et al.} 

\institute{%
  A.~Alvermann \and
  H.~Fehske \at
    \Aff{University of Greifswald,
         Institute of Physics}
        {Ernst Moritz Arndt Universit\"at Greifswald,
         Institut f\"ur Physik}
  \and
  A.~Basermann \and
  M.~R\"ohrig-Z\"ollner \and
  J.~Thies \at
    \Aff{German Aerospace Center (DLR), K\"oln}
        {Deutsches Zentrum f\"ur Luft- und Raumfahrt (DLR), K\"oln}
  \and
  H.-J.~Bungartz \and
  Th.~Huckle \and
  M.~Rippl \at
    \Aff{Technical University of Munich,
         Department of Informatics}
        {Technische Universit\"at M\"unchen,
         Fakult\"at f\"ur Informatik}
  \and
  Ch.~Carbogno \and
  M.~Scheffler \and
  D.~Simoes Brambila \at
    \Aff{Fritz Haber Institute of the Max Planck Society, Berlin}
        {Fritz-Haber-Institut der Max-Planck-Gesellschaft, Berlin}
  \and
  D.~Ernst \and
  G.~Hager \and
  M.~Kreutzer \and
  F.~Shahzad \and
  G.~Wellein \at
    \Aff{University of Erlangen-Nuremberg,
         High Performance Computing}
        {Friedrich-Alexander Universit\"at Erlangen-N\"urnberg,
         High Performance Computing}
  \and
  Y.~Futamura \and
  A.~Imakura \and
  T.~Sakurai \at
    \Aff{University of Tsukuba, Applied Mathematics}
        {University of Tsukuba, Applied Mathematics}
  \and
  M.~Galgon \and
  S.~Huber \and
  B.~Lang \and
  V.~Manin \at
    \Aff{University of Wuppertal,
         Mathematics and Natural Sciences}
        {Bergische Universit\"at Wuppertal,
         Fakult\"at f\"ur Mathematik und Naturwissenschaften},\\
      \email{lang@math.uni-wuppertal.de},
      Tel.: +49-202-4393353,
      Fax: +49-202-4392912
  \and
  A.~Ida \and
  M.~Kawai \and
  K.~Nakajima \at
    \Aff{The University of Tokyo, Computer Science}
        {The University of Tokyo, Computer Science}
  \and
  S.~K\"ocher \and
  L.~Nemec \and
  K.~Reuter \and
  Ch.~Scheurer \at
    \Aff{Technical University of Munich,
         Department of Theoretical Chemistry}
        {Technische Universit\"at M\"unchen,
         Fakult\"at f\"ur Theoretische Chemie}
  \and
  P.~Kus \and
  H.~Lederer \and
  A.~Marek \at
    \Aff{Max Planck Computing and Data Facility, Garching}
        {Max Planck Computing and Data Facility, Garching}
}

\date{Received: date / Accepted: date}

\maketitle

\begin{abstract}
We first briefly report on the status and recent achievements of the
ELPA-AEO
  (\MyEmph{E}igen\-value So\MyEmph{l}vers for
   \MyEmph{P}etaflop \MyEmph{A}pplications --
   \MyEmph{A}lgorithmic \MyEmph{E}xtensions and
   \MyEmph{O}ptimizations)
and ESSEX~II
  (\MyEmph{E}quipping \MyEmph{S}parse \MyEmph{S}olvers for
   \MyEmph{Ex}ascale)
projects.
In both collaboratory efforts, scientists from the application
areas, mathematicians, and computer scientists work together to develop
and make available efficient highly parallel methods for the solution of
eigenvalue problems.
Then we focus on a topic addressed in both projects, the use of mixed
precision computations to enhance efficiency.
We give a more detailed description of our approaches for benefiting
from either lower or higher precision in three selected contexts and of
the results thus obtained.
\end{abstract}


\section{Introduction}%
  \label{sec:Introduction}

Eigenvalue computations are at the core of simulations in various
application areas, including quantum physics and electronic structure
computations.
Being able to best utilize the capabilities of current and emerging
high-end computing systems is essential for further improving such
simulations with respect to space/time resolution or by including
additional effects in the models.
Given these needs, the ELPA-AEO and ESSEX-II projects contribute to
the development and implementation of efficient highly parallel methods
for eigenvalue problems, in different contexts.

Both projects are aimed at adding new features
(concerning, e.g., performance and resilience) to previously developed
methods and at providing additional functionality with new methods.
Building on the results of the first ESSEX funding phase
\cite{2016-KreutzerThiesEtAl-PerformanceEngineeringAnd-LNCSE113:317-338,%
      2016-ThiesGalgonEtAl-TowardsAnExascale-LNCSE113:295-316},
ESSEX-II again focuses on iterative methods for very large eigenproblems
arising, e.g., in quantum physics.
ELPA-AEO's main application area is electronic structure computation,
and for these moderately sized eigenproblems direct methods are often
superior.
Such methods are available in the widely used ELPA library
\cite{2014-MarekBlumEtAl-TheELPALibrary-JPhysCondensMatter:213201},
which had originated in an earlier project
\cite{2011-AuckenthalerBlumEtAl-ParallelSolutionOf-ParallelComput:37:783-794}
and is being improved further and extended with ELPA-AEO.

In
  Sections~\ref{sec:ELPA-AEO}
and
  \ref{sec:ESSEX}
we briefly report on the current state and on recent achievement in
the two projects, with a focus on aspects that may be of particular
interest to prospective users of the software or the underlying
methods.
In
  Section~\ref{sec:MixedPrecision}
we turn to computations involving different precisions.
Looking at three examples from the two projects we describe how lower or
higher precision is used to reduce the computing time.


\section{The ELPA-AEO project}%
  \label{sec:ELPA-AEO}

In the ELPA-AEO project, chemists, mathematicians and computer
scientists from the Max Planck Computing and Data Facility in Garching,
the Fritz Haber Institute of the Max Planck Society in Berlin, the
Technical University of Munich, and the University of Wuppertal
collaborate to provide highly scalable methods for solving
\emph{moderately-sized} ($n \lesssim 10^{6}$) Hermitian eigenvalue
problems.
Such problems arise, e.g., in electronic structure computations, and
during the earlier ELPA project, efficient direct solvers for them had
been developed and implemented in the ELPA library
  \cite{2014-MarekBlumEtAl-TheELPALibrary-JPhysCondensMatter:213201}.

This library is widely used (see
  \url{https://elpa.mpcdf.mpg.de/about}
for a description and pointers to the software), and it has been
maintained and further improved continually since the first release in
2011.
The ELPA library contains optimized routines for the steps in the direct
solution of generalized Hermitian positive eigenproblems
$A X = B X \Lambda$, that is,
(i) the Cholesky decomposition $B = U^{H} U$,
(ii) the transformation $A \mapsto \widetilde A = U^{-H} \! A U^{-1}$
 to a standard eigenproblem $\widetilde A X = \widetilde X \Lambda$,
(iii) the reduction of $\widetilde A$ to tridiagonal form, either
  in one step or via an intermediate banded matrix,
(iv) a divide-and-conquer tridiagonal eigensolver, and
(v) back-transformations for the eigenvectors corresponding to steps
  (iii) and (ii).
A typical application scenario from electronic structure computations
(``SCF cycle'') requires a sequence of a few dozens of eigenproblems
$A^{(k)} X = B X \Lambda$ to be solved, where the matrix $B$ remains
unchanged; see
  Section~\ref{sec:MixedPrecisionInELPAAEO}
for more details.
ELPA is particularly efficient in this situation by explicitly building
$U^{-1}$ for steps (ii) and (v).

ELPA-AEO is aimed at further improving the performance of computations
that are already covered by ELPA routines and at providing new
functionality.
In the remainder of this section we highlight a few recent achievements
that may be of particular interest to current and prospective users of
the library.

\begin{figure}
  \centerline{%
    \includegraphics[width=0.45\textwidth]{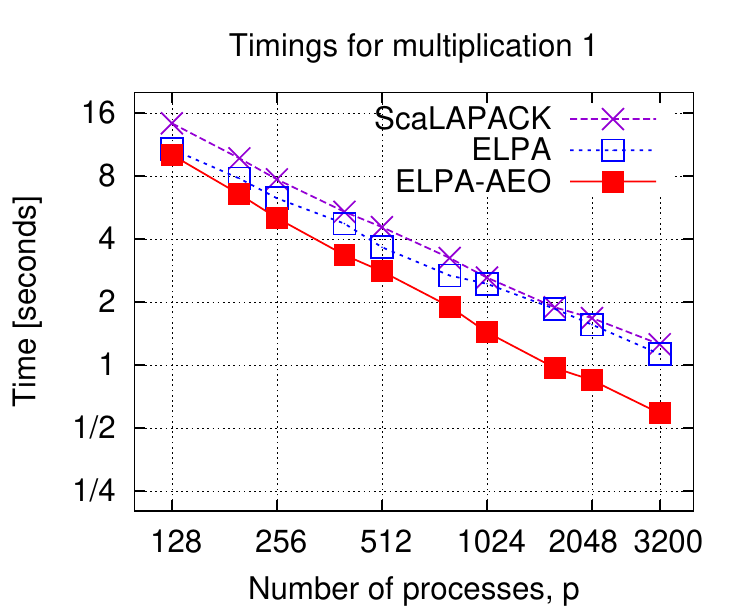}
    \quad
    \includegraphics[width=0.45\textwidth]{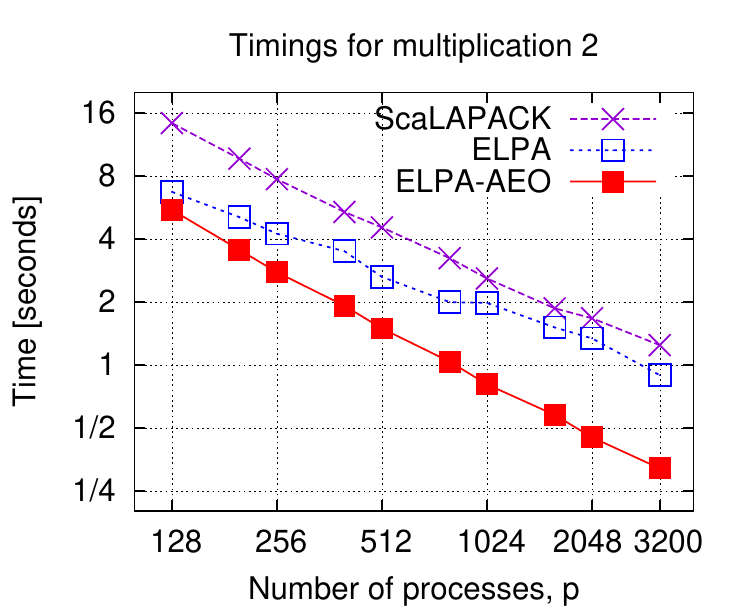}}
  \caption{Timings for the two multiplications in the transformation
           $A \mapsto \widetilde A$ with routines from ScaLAPACK,
           current ELPA routines, and the new implementations.
           The runs were made on the HYDRA system at MPCDF in Garching
           with $20$ processes per node (two $10$-core Intel Ivy
           bridge processors running at 2.8 GHz) and double precision
           real matrices of size $n = 30,000$.
           Process grids had aspect ratios $1:1$ or $1:2$; e.g., a
           $10 \times 20$ grid was set up for $p = 200$.
           With $p = 3200$, the new codes run at $\approx 40$\% of the
           nodes' \emph{peak} performance.}
  \label{fig:Cannon}
\end{figure}

An alternative approach for the transformation (ii) has been developed
  \cite{2018-ManinLang-ReductionOfGeneralized-InPreparation},
which is based on Cannon's algorithm
  \cite{1969-Cannon-ACellularComputer-PhD}.
The transformation is done with two matrix products:
\emph{multiplication~1} computes the upper triangle $M_{u}$ of
$M := A \cdot U^{-1}$,
then $M_{u}$ is transposed to obtain the lower triangle $M_{l}$ of
$M^{H} = U^{-H} A$,
and finally \emph{multiplication~2} computes the lower triangle of
$M_{l} \cdot U^{-1} = \widetilde A$.
Both routines assume that one dimension of the process grid is a
multiple of the other.
They make use of the triangular structure of their arguments
to save on computation and communication.
The timing data in
   Figure~\ref{fig:Cannon}
show that the new implementations are highly competitive.

Recent ELPA releases provide extended facilities for performance tuning.
The computational routines have an argument that can be used to guide
the routines in selecting algorithmic paths (if there are different ways
to proceed) and algorithmic parameters (such as block sizes) and to
receive performance data from their execution.
An easy-to-use autotuning facility allows setting such parameters in an
automated way by screening the parameter space; see the code fragment in
  Figure~\ref{fig:Autotune}
for an example.
Note that the parameter set obtained with the coarse probing
induced by ELPA\_AUTOTUNE\_FAST might be improved later on.

\begin{figure}
  \centerline{\ttfamily%
    \begin{tabular}{l}
      autotune\_handle = elpa\_autotune\_setup(
                           handle,
                           ELPA\_AUTOTUNE\_FAST,\\
      \In \In \In \In \In \In \In \In \In \In \In
                           ELPA\_AUTOTUNE\_DOMAIN\_REAL,
                           \&error ) ;\\
      for ( i = 0 ; i $<$ 20 ; i++ ) \{\\[0.75ex]
      \In unfinished = elpa\_autotune\_step(
                         handle, autotuning\_handle ) ;\\
      \In if ( unfinished == 0 )\\
      \In \In printf( "ELPA autotuning finished in the \%d th
                      SCF step$\backslash$n", i ) ;\\[1.5ex]
      \In /* Solve EV problem */\\
      \In elpa\_eigenvectors( handle, a, ev, z, \&error ) ;\\
     \}\\
     elpa\_autotune\_best\_set( handle, autotune\_handle ) ;\\
     elpa\_autotune\_deallocate( autotune\_handle ) ;\\
    \end{tabular}
  }
  \caption{Using ELPA's autotuning facility to adjust the algorithmic
           parameters during the solution of (at most) twenty eigenvalue
           problems in an SCF cycle, and saving them for later use.}
  \label{fig:Autotune}
\end{figure}

In earlier releases, ELPA could be configured for single or double
precision computations, but due to the naming conventions only one of
the two versions could be linked to a calling program.
Now, both precisions are accessible from one library, and mixing them
may speed up some computations; see
  Section~\ref{sec:MixedPrecisionInELPAAEO}
for an example.

New functionality for addressing banded generalized eigenvalue problems
will be added.
An efficient algorithm for the transformation to a banded standard
eigenvalue problem has been developed
  \cite{2018-Lang-EfficientReductionOf-PREPRINT},
and its parallelization is currently under way.
This will complement the functions for solving banded standard
eigenvalue problems that are already included in ELPA.


\section{The ESSEX-II project}%
  \label{sec:ESSEX}

The ESSEX-II project is a collaborative effort of physicists,
mathematicians and computer scientists from the Universities of
Erlangen-Nuremberg, Greifs\-wald, Tokyo, Tsukuba, and Wuppertal and from
the German Aerospace Center in Cologne.
It is aimed at developing exascale-enabled solvers for selected types of
\emph{very large} $(n \gg 10^{6}$) eigenproblems arising, e.g., in
quantum physics; see the project's homepage at
  \url{https://blogs.fau.de/essex/}
for more information, including pointers to publications and software.

ESSEX-II builds on results from the first ESSEX funding phase, in
particular the Exascale enabled Sparse Solver Repository (ESSR), which
provides a (block) Jacobi--Davidson method, the BEAST subspace
iteration-based framework, and the Kernel Polynomial Method (KPM) and
Chebyshev time propagation for determining few extremal eigenvalues,
a bunch of interior eigenvalues, and information about the whole
spectrum and dynamic properties, respectively.

Based on the versatile SELL-$C$-$\sigma$ format for sparse matrices
  \cite{2014-KreutzerHagerEtAl-AUnifiedSparseMatrix-SIAMJSciComput:36:C401-C423},
the \MyEmph{G}eneral, \MyEmph{H}ybrid, and \MyEmph{O}ptimized
\MyEmph{S}parse \MyEmph{T}oolkit (GHOST)
  \cite{2016-KreutzerThiesEtAl-GHOSTBuildingBlocks-IntJParallelProg:online}
contains optimized kernels for often-used operations such as sparse
matrix times (multiple) vector products (optionally fused with other
computations) and operations with block vectors, as well as a task
manager, for CPUs, Intel Xeon Phi MICs and Nvidia GPUs and
combinations of these.
The \MyEmph{P}ipelined \MyEmph{H}ybrid-parallel \MyEmph{I}terative
\MyEmph{S}olver \MyEmph{T}oolkit (PHIST)
  \cite{2016-ThiesGalgonEtAl-TowardsAnExascale-LNCSE113:295-316}
provides the eigensolver algorithms with interfaces to GHOST and other
``computational cores,'' together with higher-level functionality, such
as orthogonalization and linear solvers.

With ESSEX-II, the interoperability of these ESSR components will be
further improved to yield a mature library, which will also have an
extended range of applicability, including non-Hermitian and nonlinear
eigenproblems.
Again we highlight only a few recent achievements.

The \MyEmph{Sca}lable \MyEmph{Ma}trix \MyEmph{C}ollection (ScaMaC)
provides routines that simplify the generation of test matrices.
The matrices can be chosen from several physical models, e.g.,
boson or fermion chains, and parameters allow adjusting sizes and
physically motivated properties of the matrices.
With $32$ processes, a distributed size $2.36$G matrix for a Hubbard
model with $18$ sites and $9$ fermions can be set up in less than
$10$ minutes.

The block Jacobi--Davidson solver has been extended to non-Hermitian
and generalized eigenproblems.
It can be run with arbitrary preconditioners, e.g., the AMG
preconditioner ML
  \cite{2018-SongWubsEtAl-NumericalBifurcationAnalysis-CommunNonlinearSciNumerSimul:60:145-164},
and employs a robust and fast block orthogonalization scheme that can
make use of higher-precision computations; see
  Section~\ref{sec:HigherPrecisionForOrthogonalization}
for more details.

The BEAST framework has been extended to seamlessly integrate three
different approaches for spectral filtering in subspace iteration
methods (polynomial filters, rational filters based on plain contour
integration, and a moment-based technique) and to make use of their
respective advantages with adaptive strategies.
The BEAST framework also benefits from using different precisions; see
  Section~\ref{sec:MixedPrecisionInBeast}.

At various places, measures for improving resilience have been
included, based on verifying known properties of computed quantities and
on checksums, combined with checkpoint--restart.
To simplify incorporating the latter into numerical algorithms, the
\MyEmph{C}heckpoint--\MyEmph{R}estart and \MyEmph{A}utomatic
\MyEmph{F}ault \MyEmph{T}olerance (CRAFT) library has been developed
  \cite{2017-ShahzadThiesEtAl-CRAFTALibrary-PREPRINT}.
  Figure~\ref{fig:UsingCRAFT}
illustrates its use within the BEAST framework.
CRAFT can handle the GHOST and PHIST data types, as well as user-defined
types.
Checkpoints may be nested to accommodate, e.g., low-frequency
high-volume together with high-frequency low-volume checkpointing in
multilevel numerical algorithms, and the checkpoints can be written
asynchronously to reduce overhead.
By relying on the Scalable Checkpoint/Restart (SCR) and User-Level
Failure Mitigation (ULFM-) MPI libraries, CRAFT also provides support
for fast node-level checkpointing and for handling node failures.

\begin{figure}
  \centerline{\ttfamily%
    \begin{tabular}{l}
      // BEAST init (omitted)\\[1ex]
      Checkpoint beast\_checkpoint( "BEAST", comm ) ;\\
      beast\_checkpoint->add( "eigenvectors", \&X ) ;\\
      beast\_checkpoint->add( "eigenvalues", \&e ) ;\\
      ... // Some more\\
      beast\_checkpoint->add( "control\_variables", \&state ) ;\\
      beast\_checkpoint->commit() ;\\[1ex]
      beast\_checkpoint->restartIfNeeded( NULL ) ;\\[1ex]
      // BEAST iterations\\
      while ( !state.abort\_condition ) \{\\
      \In // Compute projector, etc. (omitted)\\
      \In ...\\
      \In beast\_checkpoint->update() ;\\
      \In beast\_checkpoint->write() ;\\
      \}
    \end{tabular}}
  \caption{Using the CRAFT library to checkpoint the current eigenvector
           approximations \texttt{X} and other quantities in every
           iteration of the main loop.}
  \label{fig:UsingCRAFT}
\end{figure}


\section{Benefits of using a different precision}%
  \label{sec:MixedPrecision}

Doing computations in lower precision is attractive from a performance
point of view because it reduces memory traffic in memory-bound code
and, in compute-bound situations, allows more operations per second,
due to vector instructions manipulating more elements at a time.
However, the desired accuracy often cannot be reached in single
precision and then only a part of the computations can be done in
lower precision, or a correction is needed; cf., e.g.,
  \cite{baboulin2009accelerating}
for the latter.
In
  Subsection~\ref{sec:MixedPrecisionInBeast}
we describe an approach for reducing overall runtimes of the BEAST
framework by using lower-precision computations for early iterations.

Higher precision, on the other hand, is often a means to improve
robustness.
It is less known that higher precision can also be beneficial w.r.t.\
runtime.
This is demonstrated in
  Subsection~\ref{sec:HigherPrecisionForOrthogonalization}
in the context of orthogonalization.

In
  Section~\ref{sec:MixedPrecisionInELPAAEO}
we come back to using lower precision, from the perspective of an
important application area: self-consistent field (SCF) cycles in
electronic structure computations.
Each iteration of such a cycle requires the solution of a generalized
eigenproblem (GEP).
After briefly introducing the context, we discuss how ELPA-AEO's
features can be used to steer the precision from the application code,
targeting either the entire solution of a GEP or particular steps within
its solution.


\subsection{Changing precision in subspace iteration-based eigensolvers}
  \label{sec:MixedPrecisionInBeast}

The BEAST framework
  \cite{2017-GalgonKraemerLang-ImprovingProjectionBased-NumerLinearAlgebraAppl:2017:e2124,%
        2016-ThiesGalgonEtAl-TowardsAnExascale-LNCSE113:295-316}
is aimed at finding those eigenpairs $(\lambda, x)$ of a generalized
interior eigenproblem $A x = B x \lambda$ ($A$ Hermitian, $B$ Hermitian
positive definite) with $\lambda$ in a given search interval
$I_{\lambda} = [ \underline{\lambda}, \overline{\lambda} ]$, in
particular for interior eigenvalues.
It is based on subspace iteration with spectral filtering and
Rayleigh--Ritz extraction, that is, a subspace $U$ containing an
approximate basis for the desired eigenvectors is constructed from some
initial vectors $Y$, then a Rayleigh--Ritz step is used to obtain the
approximate eigenpairs.
If the desired residual threshold is not yet reached, we iterate, using
the approximate eigenvectors in our choice of $Y$ for the following
iteration; cf.\ also
  Figure~\ref{fig:beast}
below.
The main distinguishing factor of the variants BEAST-P/-C/-M in our
framework is the construction of the subspace $U$.

BEAST-P, which is only applicable for standard eigenproblems, implements
a polynomial filter
  \cite{2016-PieperKreutzerEtAl-HighPerformanceImplementation-JComputPhys:325:226-243,%
        doi:10.1137/1.9781611970739}, 
using matrix--(block) vector products to apply a polynomial in $A$ to
$Y$,
\[
  U = \sum\limits_{j=0}^{N}{ \omega_{j} A^{j} Y }
  .
\]

In both BEAST-C and BEAST-M, the filter is applied via quadrature
approximations of contour integrals of the form
\[
  r( B^{-1} A )
  \approx
  \frac{1}{2 \pi i} \int_{\Gamma} z^{k} ( z B - A )^{-1} B \dz
  ,
\]
where $\Gamma$ is a contour in the complex plane enclosing the sought
eigenvalues and no others.
BEAST-C follows Polizzi's FEAST algorithm
  \cite{polizzi2009density}
in computing
\[
  U = \sum_{j=1}^{N} w_{j} ( z_{j} B - A )^{-1} B Y
\]
with suitable nodes $z_{j}$ and weights $w_{j}$.
This requires $N$ linear solves for each iteration of the eigensolver,
with an $n \times m$ block vector of right hand sides $Y$.
BEAST-M realizes a specific Sakurai--Sugiura method
  \cite{sakurai2003projection},
Sakurai--Sugiura Rayleigh--Ritz
  \cite{sakurai2007cirr}.
Here, the subspace is constructed as
\[
  U = \left[ U_{0}, ..., U_{s-1} \right]
  , \quad \mbox{where} \quad
  U_{k} = \sum_{j=1}^{N} w_{j} z_{j}^{k} ( z_{j} B - A )^{-1} B Y
  .
\]
Thus, again $N$ linear solves must be performed as in BEAST-C, but since
the overall subspace is computed as a combination of their solution, $Y$
needs only $(1/s)$th the desired number of columns of $U$, which can
reduce the cost of the linear solves.
It should be noted that a traditional Sakurai--Sugiura Rayleigh--Ritz
implementation requires very few, or only one iteration, with a large
overall subspace size.
However, we consider it here within the context of a constrained
subspace size, making it a truly iterative method.

\begin{figure}
  \centerline{%
    \includegraphics[width=0.5\linewidth]{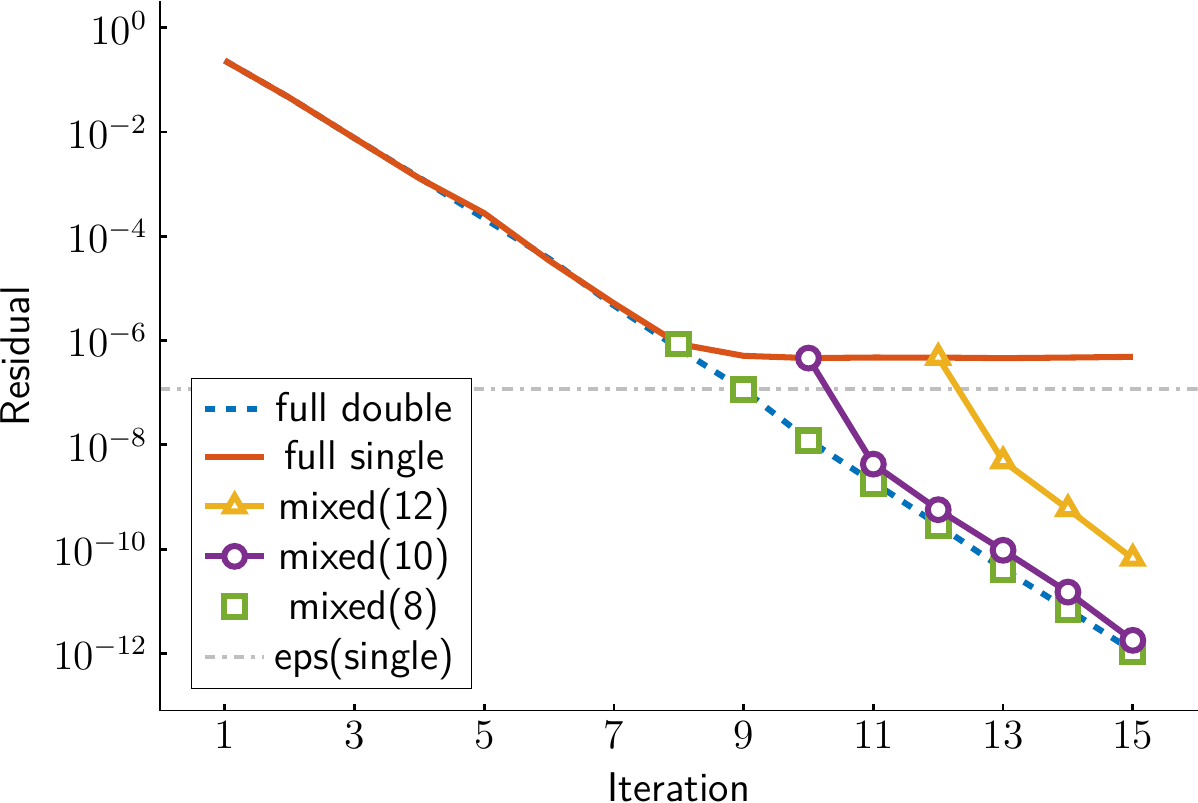}}
  \caption{Smallest residual
           $\min_{i} \| A x_{i} - \lambda_{i} x_{i} \|$
           over BEAST-P iterations (with polynomial degree $50$) for
           computations done completely in double precision, completely
           in single precision, and with switching from single to double
           precision after iterations $12$, $10$, and $8$.
           The horizontal line indicates the single precision machine
           epsilon.}
  \label{fig:convergencepoly}
\end{figure}

We first consider the effect of starting with single precision and
switching to double precision in later iterations.
Since BEAST is designed to behave iteratively, we expect that this
effect should be limited. 
  Figure~\ref{fig:convergencepoly}
shows BEAST-P's progress (smallest residual of the current
approximations in each iteration) in solving a standard eigenproblem
$A X = X \Lambda$ for a size $3200$ topological insulator matrix $A$
from the ESSEX repository
  \cite{2014-AlvermannBasermannEtAl-ESSEXEquippingSparse-ProcEuroPar2014-LNCS:577-588}
and $I_{\lambda} = [ -0.5, 0.5 ]$, which contains $36$ eigenpairs.
We see that the residuals for single precision data and computations are
very close to those obtained with double precision, until we reach the
single precision barrier.
Continuing in single precision leads to stagnation.
By contrast, if we switch to double precision data and computations
sufficiently before the barrier, convergence proceeds as if the entire
run was in double precision.
Even a later switch need not have dramatic effects; we see that
convergence, although stalled temporarily by the single precision
barrier, proceeds at the same rate and possibly even slightly faster
when switched two and four iterations ``too late.''
In the case of $10$ iterations in single precision (two past the ideal
of $8$), the overall residual reached after $15$ total iterations is
again close to that of the full double and ideal switch computations.

\begin{figure}
  \centerline{%
    \includegraphics[width=0.85\linewidth]{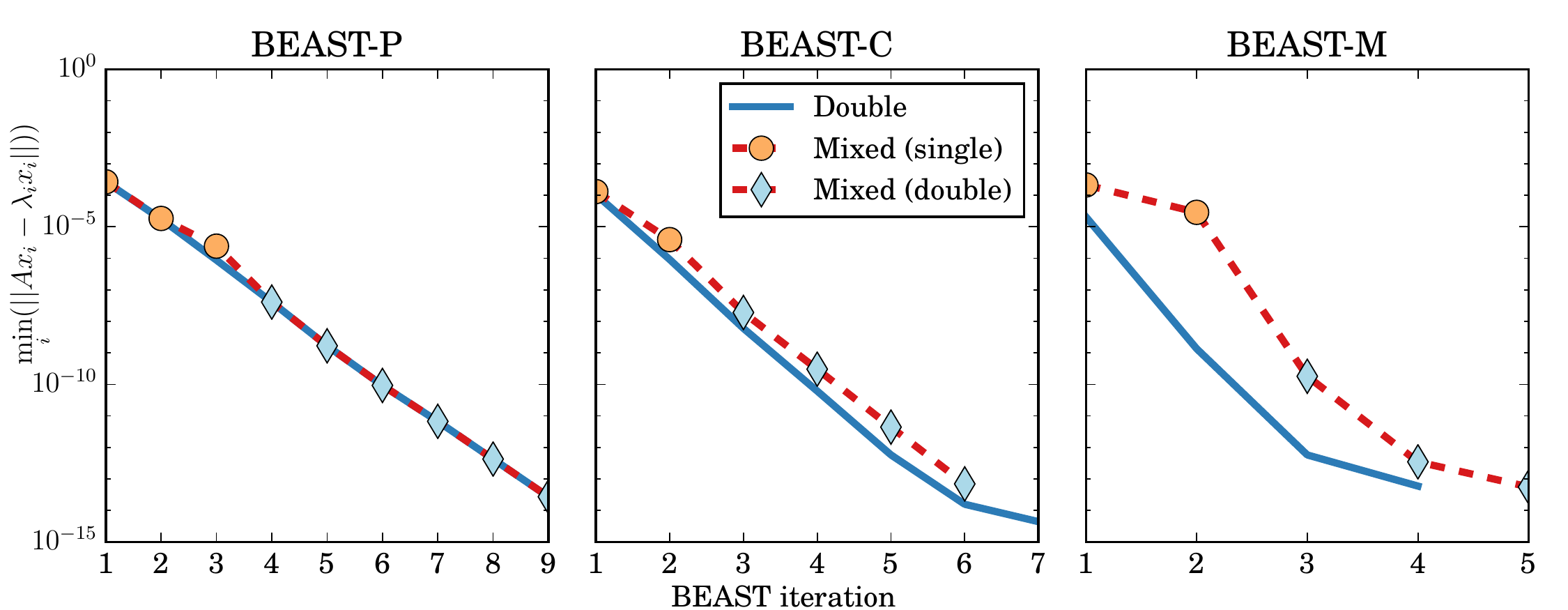}}
  \caption{Smallest residual over BEAST iterations for runs done
           completely in double precision (solid lines) and for
           mixed precision runs (dashed lines; different markers for
           iterations done in single and double precision,
           respectively), with algorithmic parameters set as follows.
           Size of $U$: $1.5 \times$ the estimated number of eigenvalues
           in the interval;
           polynomial degree of BEAST-P: $10,\!000$;
           $4$ Gauss-Legendre integration points for BEAST-C and $8$ for
           BEAST-M, which is more sensitive to a low number of nodes;
           STRUMPACK direct solver
             \cite{Rouet:2016:DPD:2956571.2930660}
           for the linear systems in BEAST-C and -M;
           threshold for the switch from single to double precision:
           $10^{-5}$ for BEAST-P and -C, and $10^{-4}$ for BEAST-M to
           prevent excessive stagnation.}
  \label{fig:convergence}
\end{figure}

A switching strategy based on this observation is shown in
  Figure~\ref{fig:beast}.
In
  Figure~\ref{fig:convergence}
we report results for using this approach to solve the problem
$A X = \Lambda X$ for a size $16$M graphene matrix from the ESSEX
repository and $I_{\lambda} =  \left[ -0.0025, 0.0025 \right]$.
The computation was done on the Emmy cluster at the University of
Erlangen-Nuremberg, using 32 nodes, each with two Xeon 2660v2 chips.
All methods computed an identical number of 318 eigenpairs in
$I_{\lambda}$ to a tolerance of $10^{-10}$.
BEAST-P exhibits a remarkable similarity in convergence rates between
single and double precision before the switch threshold, and the mixed
precision run was roughly $1.2$ times faster than using double precision
throughout.
In BEAST-C the rates are again similar; due to a few unconverged
eigenpairs, the double precision computation required an additional
iteration of the eigensolver for this problem, enabling a higher speedup
$1.4$ for the mixed precision version.
In BEAST-M, we observe some stagnation before the switch threshold, and
an additional iteration was required in the mixed precision run.
In this case, the mixed precision run was slower than pure double
precision, with a ``speedup'' of $0.9$.
Overall, the reduction in time from early iterations performed in single
precision shows most clearly for BEAST-P.
We note that the actual speed-up observed between single and double
precision depends on both the hardware and software used; higher
optimization of vectorized instructions or the use of accelerators such
as GPUs could produce a more dramatic time difference.

\begin{figure}
  \qquad
  \parbox{0.92\textwidth}{%
    \SetAlgoLined
    Choose desired subspace size $m$
      ($> \mbox{number of evals in $I_{\lambda}$}$)
      and initial vectors $Y$ \\
    \While{not converged}{
      Construct subspace $U \leftarrow Y$ with BEAST-* scheme\\
      Resize subspace based on $\mathrm{rank}( U )$\\
      Solve reduced eigenproblem $A_U W = B_U W \Lambda$, where
        $A_U = U^{*} A U$, $B_U = U^{*} B U$ \\
      $X := U W$ \\
      $Y := B X$ (BEAST-P/-C) or
        $Y := B X R$ (BEAST-M, with a random matrix $R$) \\
      If single precision barrier has been reached, switch to double
        precision
    }
  }
  \caption{The mixed-precision BEAST framework.
           Computations are started in single precision and may be
           continued in double precision.}
  \label{fig:beast}
\end{figure} 

The results indicate that initial iterations in single precision may
have a limited effect on the overall convergence of the eigensolver if
an appropriate switching point to double precision is chosen, thus
allowing for a reduction in cost without sacrificing accuracy.
We plan to combine this approach with relaxed stopping criteria for
solving the linear systems in BEAST-C and -M iteratively; cf.\ also
  \cite{2017-GalgonKraemerLang-ImprovingProjectionBased-NumerLinearAlgebraAppl:2017:e2124,%
        doi:10.1002/nla.2188}
for related work.


\subsection{Using higher precision for robust and fast
            orthogonalization}%
  \label{sec:HigherPrecisionForOrthogonalization}

In contrast to the standard Jacobi--Davidson method, which determines
the sought eigenpairs one-by-one, the block Jacobi--Davison method in
ESSEX
  \cite{2015-RoehrigZoellnerThiesEtAl-IncreasingThePerformance-SIAMJSciComput:37:C697-C722}
computes them by groups.
Here we will consider only the real standard eigenvalue problem
$A v_{i} = v_{i} \lambda_{i}$.
Then one iteration contains the following two major steps:
\begin{enumerate}
  \item Given $n_{b}$ current approximations $\tilde \lambda_{i}$ and
    $\tilde v_{i}$, $i = 1, \ldots, n_{b}$, and a set of previously
    converged Schur vectors $W = ( w_{1}, \ldots, w_{k} )$ ($k \geq 0$),
    use some steps of a (blocked) iterative linear solver for the
    correction equation
    \[
        ( I - \tilde W \tilde W^{T} )
        ( A - \tilde \lambda_{i} I )
        ( I - \tilde W \tilde W^{T} )
        x_{i}
      \; = \;
        - r_{i},
      \quad
      i = 1, \ldots, n_{b},
    \]
    where $r_{i} = A \tilde v_{i} - \tilde v_{i} \tilde \lambda_{i}$ are
    the current residuals and
    $\tilde W = ( W \; | \; \tilde v_{1}, \ldots, \tilde v_{n_{b}} )$.
  \item Obtain new directions $y_{1}, \ldots, y_{n_{b}}$ by
    orthogonalizing the $x_{i}$ against $W$ and among themselves.
    (The $y_{i}$ are then used to update the $\tilde v_{i}$.)
\end{enumerate}

The block method typically requires more operations than the non-blocked
one and therefore has previously not been advocated, but in
  \cite{2015-RoehrigZoellnerThiesEtAl-IncreasingThePerformance-SIAMJSciComput:37:C697-C722}
it has been shown that this drawback can be more than outweighed by
allowing the use of kernels that can be implemented to make best use of
the capabilities of modern processors (in particular, sparse matrix
times multiple vectors), such that the block method tends to run faster.
In addition, it is more robust in the presence of multiple or tightly
clustered eigenvalues.

In the following we focus on the orthogonalization in step~2.
It is well known that if one first orthogonalizes the $x_{i}$ against
$W$ (``\emph{phase~I}'') and then among themselves
(``\emph{phase~II}''), the second phase can spoil the results of the
first one; this also holds if we reverse the order of the phases.
By contrast, a robust algorithm is obtained by iterating this process,
alternating between the two phases and using a rank-revealing technique
in phase~II; see
  \cite{Hoemmen2010thesis,%
        Stewart2008bgs}
for a thorough discussion.

We follow this approach, using a plain projection
$\tilde Y = ( I - W W^{T} ) X$ for phase~I
and SVQB
  \cite{Stathopoulos2002}
on $\tilde Y$ for phase~II.
We prefer SVQB over TSQR
  \cite{Demmel2012}
because the bulk of computation may be done in a highly
performant matrix--matrix multiplication for building the Gram matrix
$M = \tilde Y^{T} \tilde Y$.
This would also be true for CholQR
  \cite{Stathopoulos2002},
but SVQB is superior in the following sense.

Both methods orthogonalize $\tilde Y$ by determining a suitable matrix
$Z \in \R^{n_{b} \times n_{b}}$ such that $Z^{T} M Z = I$, and setting
$Y = \tilde Y Z$; this yields $Y^{T} Y = I$.
For SVQB we take $Z = U \Lambda^{-1/2}$, where $M = U \Lambda U^{T}$ is
an eigendecomposition of $M$,
whereas a (possibly pivoted, partial) Cholesky decomposition
$M = R^{T} R$ is used for setting $Z = R^{-1}$ in CholQR.
In order to minimize amplification of rounding errors in the final
multiplication $\tilde Y Z$, $Z$ should be as close as possible to the
identity matrix, that is, we have to solve the optimization problem
\[
  \min_{Z \in \R^{n_{b} \times n_{b}}, \, Z^{T} M Z = I}
    \| Z - I \|
  .
\]
For the Frobenius norm, this is a special case of the orthogonal
Procrustes problem analyzed by Sch\"onemann in
  \cite{Schoenemann1966},
as it can be transformed to the following formulation:
\[
  \min_{\hat Z \in \R^{n_{b} \times n_{b}}, \, \hat Z^{T} \hat Z = I}
    \| \bar M^{1/2} \hat Z - I \|_{F}
  .
\]
As shown in
  \cite{Schoenemann1966},
a solution can be constructed as $\hat Z = U U^{T}$ with the
eigendecomposition $M^{1/2} = U D^{1/2} U^{T}$ (in
  \cite{Schoenemann1966}
a more general case is considered exploiting a singular value
decomposition).
So the choice $Z = M^{-1/2} \hat Z = U D^{-1/2}$, that is, the SVQB
algorithm, is optimal in the sense discussed above.
(For simplicity of the presentation we have assumed $M$ to be full-rank,
thus symmetric positive definite.
The argumentation also can be extended to the rank-deficient case.)

Our aim is to obtain a robust and fast overall orthogonalization
method with fewer iterations by using extended precision computations;
cf.\ 
  \cite{ytdd14vecpar,ytd15sisc}
for related ideas in the context of CholQR and communication-avoiding
GMRES.

In contrast to
  \cite{ytdd14vecpar,ytd15sisc},
we use extended precision throughout the orthogonalization,
including the orthogonalization against $W$ and the computation and
decomposition of the Gram matrix.
Our own kernels are based on the techniques described in
  \cite{Muller2010}
for working with numbers represented by two doubles (DD).
Some of the kernels take standard double precision data and return
DD results, others also take DD data as inputs.
They make use of AVX2, Intel's \emph{advanced vector extensions}, with
FMA (fused multiply--add) operations; see
  \cite[Chapter 5]{Muller2010}.
As proposed there, divisions and square roots are computed using the
Newton--Raphson method.

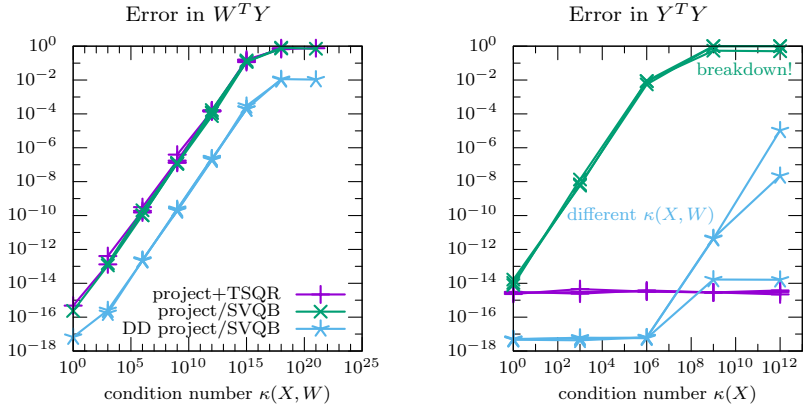
\begin{figure}
  \begin{center}
    \begin{tabular}{c@{\qquad}c}
      Error in $W^{T} Y$ & Error in $Y^{T} Y$ \\
      \begin{tikzpicture}[gnuplot]
\tikzset{every node/.append style={scale=0.85}}
\path (0.000,0.000) rectangle (5.000,5.000);
\gpcolor{color=gp lt color border}
\gpsetlinetype{gp lt border}
\gpsetlinewidth{1.00}
\draw[gp path] (0.855,0.729)--(1.035,0.729);
\draw[gp path] (4.657,0.729)--(4.477,0.729);
\node[gp node right] at (0.741,0.729) {$10^{-18}$};
\draw[gp path] (0.855,0.954)--(0.945,0.954);
\draw[gp path] (4.657,0.954)--(4.567,0.954);
\draw[gp path] (0.855,1.178)--(1.035,1.178);
\draw[gp path] (4.657,1.178)--(4.477,1.178);
\node[gp node right] at (0.741,1.178) {$10^{-16}$};
\draw[gp path] (0.855,1.403)--(0.945,1.403);
\draw[gp path] (4.657,1.403)--(4.567,1.403);
\draw[gp path] (0.855,1.627)--(1.035,1.627);
\draw[gp path] (4.657,1.627)--(4.477,1.627);
\node[gp node right] at (0.741,1.627) {$10^{-14}$};
\draw[gp path] (0.855,1.852)--(0.945,1.852);
\draw[gp path] (4.657,1.852)--(4.567,1.852);
\draw[gp path] (0.855,2.076)--(1.035,2.076);
\draw[gp path] (4.657,2.076)--(4.477,2.076);
\node[gp node right] at (0.741,2.076) {$10^{-12}$};
\draw[gp path] (0.855,2.301)--(0.945,2.301);
\draw[gp path] (4.657,2.301)--(4.567,2.301);
\draw[gp path] (0.855,2.525)--(1.035,2.525);
\draw[gp path] (4.657,2.525)--(4.477,2.525);
\node[gp node right] at (0.741,2.525) {$10^{-10}$};
\draw[gp path] (0.855,2.750)--(0.945,2.750);
\draw[gp path] (4.657,2.750)--(4.567,2.750);
\draw[gp path] (0.855,2.975)--(1.035,2.975);
\draw[gp path] (4.657,2.975)--(4.477,2.975);
\node[gp node right] at (0.741,2.975) {$10^{-8}$};
\draw[gp path] (0.855,3.199)--(0.945,3.199);
\draw[gp path] (4.657,3.199)--(4.567,3.199);
\draw[gp path] (0.855,3.424)--(1.035,3.424);
\draw[gp path] (4.657,3.424)--(4.477,3.424);
\node[gp node right] at (0.741,3.424) {$10^{-6}$};
\draw[gp path] (0.855,3.648)--(0.945,3.648);
\draw[gp path] (4.657,3.648)--(4.567,3.648);
\draw[gp path] (0.855,3.873)--(1.035,3.873);
\draw[gp path] (4.657,3.873)--(4.477,3.873);
\node[gp node right] at (0.741,3.873) {$10^{-4}$};
\draw[gp path] (0.855,4.097)--(0.945,4.097);
\draw[gp path] (4.657,4.097)--(4.567,4.097);
\draw[gp path] (0.855,4.322)--(1.035,4.322);
\draw[gp path] (4.657,4.322)--(4.477,4.322);
\node[gp node right] at (0.741,4.322) {$10^{-2}$};
\draw[gp path] (0.855,4.546)--(0.945,4.546);
\draw[gp path] (4.657,4.546)--(4.567,4.546);
\draw[gp path] (0.855,4.771)--(1.035,4.771);
\draw[gp path] (4.657,4.771)--(4.477,4.771);
\node[gp node right] at (0.741,4.771) {$10^{0}$};
\draw[gp path] (0.855,0.729)--(0.855,0.909);
\draw[gp path] (0.855,4.771)--(0.855,4.591);
\node[gp node center] at (0.855,0.501) {$10^{0}$};
\draw[gp path] (1.007,0.729)--(1.007,0.819);
\draw[gp path] (1.007,4.771)--(1.007,4.681);
\draw[gp path] (1.159,0.729)--(1.159,0.819);
\draw[gp path] (1.159,4.771)--(1.159,4.681);
\draw[gp path] (1.311,0.729)--(1.311,0.819);
\draw[gp path] (1.311,4.771)--(1.311,4.681);
\draw[gp path] (1.463,0.729)--(1.463,0.819);
\draw[gp path] (1.463,4.771)--(1.463,4.681);
\draw[gp path] (1.615,0.729)--(1.615,0.909);
\draw[gp path] (1.615,4.771)--(1.615,4.591);
\node[gp node center] at (1.615,0.501) {$10^{5}$};
\draw[gp path] (1.767,0.729)--(1.767,0.819);
\draw[gp path] (1.767,4.771)--(1.767,4.681);
\draw[gp path] (1.920,0.729)--(1.920,0.819);
\draw[gp path] (1.920,4.771)--(1.920,4.681);
\draw[gp path] (2.072,0.729)--(2.072,0.819);
\draw[gp path] (2.072,4.771)--(2.072,4.681);
\draw[gp path] (2.224,0.729)--(2.224,0.819);
\draw[gp path] (2.224,4.771)--(2.224,4.681);
\draw[gp path] (2.376,0.729)--(2.376,0.909);
\draw[gp path] (2.376,4.771)--(2.376,4.591);
\node[gp node center] at (2.376,0.501) {$10^{10}$};
\draw[gp path] (2.528,0.729)--(2.528,0.819);
\draw[gp path] (2.528,4.771)--(2.528,4.681);
\draw[gp path] (2.680,0.729)--(2.680,0.819);
\draw[gp path] (2.680,4.771)--(2.680,4.681);
\draw[gp path] (2.832,0.729)--(2.832,0.819);
\draw[gp path] (2.832,4.771)--(2.832,4.681);
\draw[gp path] (2.984,0.729)--(2.984,0.819);
\draw[gp path] (2.984,4.771)--(2.984,4.681);
\draw[gp path] (3.136,0.729)--(3.136,0.909);
\draw[gp path] (3.136,4.771)--(3.136,4.591);
\node[gp node center] at (3.136,0.501) {$10^{15}$};
\draw[gp path] (3.288,0.729)--(3.288,0.819);
\draw[gp path] (3.288,4.771)--(3.288,4.681);
\draw[gp path] (3.440,0.729)--(3.440,0.819);
\draw[gp path] (3.440,4.771)--(3.440,4.681);
\draw[gp path] (3.592,0.729)--(3.592,0.819);
\draw[gp path] (3.592,4.771)--(3.592,4.681);
\draw[gp path] (3.745,0.729)--(3.745,0.819);
\draw[gp path] (3.745,4.771)--(3.745,4.681);
\draw[gp path] (3.897,0.729)--(3.897,0.909);
\draw[gp path] (3.897,4.771)--(3.897,4.591);
\node[gp node center] at (3.897,0.501) {$10^{20}$};
\draw[gp path] (4.049,0.729)--(4.049,0.819);
\draw[gp path] (4.049,4.771)--(4.049,4.681);
\draw[gp path] (4.201,0.729)--(4.201,0.819);
\draw[gp path] (4.201,4.771)--(4.201,4.681);
\draw[gp path] (4.353,0.729)--(4.353,0.819);
\draw[gp path] (4.353,4.771)--(4.353,4.681);
\draw[gp path] (4.505,0.729)--(4.505,0.819);
\draw[gp path] (4.505,4.771)--(4.505,4.681);
\draw[gp path] (4.657,0.729)--(4.657,0.909);
\draw[gp path] (4.657,4.771)--(4.657,4.591);
\node[gp node center] at (4.657,0.501) {$10^{25}$};
\draw[gp path] (0.855,4.771)--(0.855,0.729)--(4.657,0.729)--(4.657,4.771)--cycle;
\node[gp node center] at (2.756,0.159) {condition number $\kappa(X,W)$};
\node[gp node right] at (3.679,1.479) {project+TSQR};
\gpcolor{rgb color={0.580,0.000,0.827}}
\gpsetlinewidth{2.00}
\draw[gp path] (3.793,1.479)--(4.429,1.479);
\draw[gp path] (0.855,1.331)--(1.311,1.878)--(1.767,2.566)--(2.224,3.233)--(2.680,3.918);
\draw[gp path] (1.311,1.988)--(1.767,2.587)--(2.224,3.224)--(2.680,3.920)--(3.136,4.560);
\draw[gp path] (1.767,2.637)--(2.224,3.254)--(2.680,3.908)--(3.136,4.576)--(3.592,4.744);
\draw[gp path] (2.224,3.333)--(2.680,3.906)--(3.136,4.593)--(3.592,4.732)--(4.049,4.744);
\gpsetpointsize{8.00}
\gppoint{gp mark 1}{(0.855,1.331)}
\gppoint{gp mark 1}{(1.311,1.878)}
\gppoint{gp mark 1}{(1.767,2.566)}
\gppoint{gp mark 1}{(2.224,3.233)}
\gppoint{gp mark 1}{(2.680,3.918)}
\gppoint{gp mark 1}{(1.311,1.988)}
\gppoint{gp mark 1}{(1.767,2.587)}
\gppoint{gp mark 1}{(2.224,3.224)}
\gppoint{gp mark 1}{(2.680,3.920)}
\gppoint{gp mark 1}{(3.136,4.560)}
\gppoint{gp mark 1}{(1.767,2.637)}
\gppoint{gp mark 1}{(2.224,3.254)}
\gppoint{gp mark 1}{(2.680,3.908)}
\gppoint{gp mark 1}{(3.136,4.576)}
\gppoint{gp mark 1}{(3.592,4.744)}
\gppoint{gp mark 1}{(2.224,3.333)}
\gppoint{gp mark 1}{(2.680,3.906)}
\gppoint{gp mark 1}{(3.136,4.593)}
\gppoint{gp mark 1}{(3.592,4.732)}
\gppoint{gp mark 1}{(4.049,4.744)}
\gppoint{gp mark 1}{(4.111,1.479)}
\gpcolor{color=gp lt color border}
\node[gp node right] at (3.679,1.251) {project/SVQB};
\gpcolor{rgb color={0.000,0.620,0.451}}
\draw[gp path] (3.793,1.251)--(4.429,1.251);
\draw[gp path] (0.855,1.261)--(1.311,1.864)--(1.767,2.530)--(2.224,3.222)--(2.680,3.884);
\draw[gp path] (1.311,1.893)--(1.767,2.595)--(2.224,3.217)--(2.680,3.924)--(3.136,4.582);
\draw[gp path] (1.767,2.528)--(2.224,3.218)--(2.680,3.893)--(3.136,4.563)--(3.592,4.740);
\draw[gp path] (2.224,3.209)--(2.680,3.844)--(3.136,4.565)--(3.592,4.751)--(4.049,4.737);
\gppoint{gp mark 2}{(0.855,1.261)}
\gppoint{gp mark 2}{(1.311,1.864)}
\gppoint{gp mark 2}{(1.767,2.530)}
\gppoint{gp mark 2}{(2.224,3.222)}
\gppoint{gp mark 2}{(2.680,3.884)}
\gppoint{gp mark 2}{(1.311,1.893)}
\gppoint{gp mark 2}{(1.767,2.595)}
\gppoint{gp mark 2}{(2.224,3.217)}
\gppoint{gp mark 2}{(2.680,3.924)}
\gppoint{gp mark 2}{(3.136,4.582)}
\gppoint{gp mark 2}{(1.767,2.528)}
\gppoint{gp mark 2}{(2.224,3.218)}
\gppoint{gp mark 2}{(2.680,3.893)}
\gppoint{gp mark 2}{(3.136,4.563)}
\gppoint{gp mark 2}{(3.592,4.740)}
\gppoint{gp mark 2}{(2.224,3.209)}
\gppoint{gp mark 2}{(2.680,3.844)}
\gppoint{gp mark 2}{(3.136,4.565)}
\gppoint{gp mark 2}{(3.592,4.751)}
\gppoint{gp mark 2}{(4.049,4.737)}
\gppoint{gp mark 2}{(4.111,1.251)}
\gpcolor{color=gp lt color border}
\node[gp node right] at (3.679,1.023) {DD project/SVQB};
\gpcolor{rgb color={0.337,0.706,0.914}}
\draw[gp path] (3.793,1.023)--(4.429,1.023);
\draw[gp path] (0.855,0.914)--(1.311,1.270)--(1.767,1.926)--(2.224,2.611)--(2.680,3.283);
\draw[gp path] (1.311,1.231)--(1.767,1.930)--(2.224,2.600)--(2.680,3.280)--(3.136,3.937);
\draw[gp path] (1.767,1.938)--(2.224,2.586)--(2.680,3.259)--(3.136,3.979)--(3.592,4.324);
\draw[gp path] (2.224,2.592)--(2.680,3.269)--(3.136,3.939)--(3.592,4.334)--(4.049,4.328);
\gppoint{gp mark 3}{(0.855,0.914)}
\gppoint{gp mark 3}{(1.311,1.270)}
\gppoint{gp mark 3}{(1.767,1.926)}
\gppoint{gp mark 3}{(2.224,2.611)}
\gppoint{gp mark 3}{(2.680,3.283)}
\gppoint{gp mark 3}{(1.311,1.231)}
\gppoint{gp mark 3}{(1.767,1.930)}
\gppoint{gp mark 3}{(2.224,2.600)}
\gppoint{gp mark 3}{(2.680,3.280)}
\gppoint{gp mark 3}{(3.136,3.937)}
\gppoint{gp mark 3}{(1.767,1.938)}
\gppoint{gp mark 3}{(2.224,2.586)}
\gppoint{gp mark 3}{(2.680,3.259)}
\gppoint{gp mark 3}{(3.136,3.979)}
\gppoint{gp mark 3}{(3.592,4.324)}
\gppoint{gp mark 3}{(2.224,2.592)}
\gppoint{gp mark 3}{(2.680,3.269)}
\gppoint{gp mark 3}{(3.136,3.939)}
\gppoint{gp mark 3}{(3.592,4.334)}
\gppoint{gp mark 3}{(4.049,4.328)}
\gppoint{gp mark 3}{(4.111,1.023)}
\gpcolor{color=gp lt color border}
\gpsetlinewidth{1.00}
\draw[gp path] (0.855,4.771)--(0.855,0.729)--(4.657,0.729)--(4.657,4.771)--cycle;
\gpdefrectangularnode{gp plot 1}{\pgfpoint{0.855cm}{0.729cm}}{\pgfpoint{4.657cm}{4.771cm}}
\end{tikzpicture}
 & \begin{tikzpicture}[gnuplot]
\tikzset{every node/.append style={scale=0.85}}
\path (0.000,0.000) rectangle (5.000,5.000);
\gpcolor{color=gp lt color border}
\gpsetlinetype{gp lt border}
\gpsetlinewidth{1.00}
\draw[gp path] (0.855,0.729)--(1.035,0.729);
\draw[gp path] (4.657,0.729)--(4.477,0.729);
\node[gp node right] at (0.741,0.729) {$10^{-18}$};
\draw[gp path] (0.855,0.954)--(0.945,0.954);
\draw[gp path] (4.657,0.954)--(4.567,0.954);
\draw[gp path] (0.855,1.178)--(1.035,1.178);
\draw[gp path] (4.657,1.178)--(4.477,1.178);
\node[gp node right] at (0.741,1.178) {$10^{-16}$};
\draw[gp path] (0.855,1.403)--(0.945,1.403);
\draw[gp path] (4.657,1.403)--(4.567,1.403);
\draw[gp path] (0.855,1.627)--(1.035,1.627);
\draw[gp path] (4.657,1.627)--(4.477,1.627);
\node[gp node right] at (0.741,1.627) {$10^{-14}$};
\draw[gp path] (0.855,1.852)--(0.945,1.852);
\draw[gp path] (4.657,1.852)--(4.567,1.852);
\draw[gp path] (0.855,2.076)--(1.035,2.076);
\draw[gp path] (4.657,2.076)--(4.477,2.076);
\node[gp node right] at (0.741,2.076) {$10^{-12}$};
\draw[gp path] (0.855,2.301)--(0.945,2.301);
\draw[gp path] (4.657,2.301)--(4.567,2.301);
\draw[gp path] (0.855,2.525)--(1.035,2.525);
\draw[gp path] (4.657,2.525)--(4.477,2.525);
\node[gp node right] at (0.741,2.525) {$10^{-10}$};
\draw[gp path] (0.855,2.750)--(0.945,2.750);
\draw[gp path] (4.657,2.750)--(4.567,2.750);
\draw[gp path] (0.855,2.975)--(1.035,2.975);
\draw[gp path] (4.657,2.975)--(4.477,2.975);
\node[gp node right] at (0.741,2.975) {$10^{-8}$};
\draw[gp path] (0.855,3.199)--(0.945,3.199);
\draw[gp path] (4.657,3.199)--(4.567,3.199);
\draw[gp path] (0.855,3.424)--(1.035,3.424);
\draw[gp path] (4.657,3.424)--(4.477,3.424);
\node[gp node right] at (0.741,3.424) {$10^{-6}$};
\draw[gp path] (0.855,3.648)--(0.945,3.648);
\draw[gp path] (4.657,3.648)--(4.567,3.648);
\draw[gp path] (0.855,3.873)--(1.035,3.873);
\draw[gp path] (4.657,3.873)--(4.477,3.873);
\node[gp node right] at (0.741,3.873) {$10^{-4}$};
\draw[gp path] (0.855,4.097)--(0.945,4.097);
\draw[gp path] (4.657,4.097)--(4.567,4.097);
\draw[gp path] (0.855,4.322)--(1.035,4.322);
\draw[gp path] (4.657,4.322)--(4.477,4.322);
\node[gp node right] at (0.741,4.322) {$10^{-2}$};
\draw[gp path] (0.855,4.546)--(0.945,4.546);
\draw[gp path] (4.657,4.546)--(4.567,4.546);
\draw[gp path] (0.855,4.771)--(1.035,4.771);
\draw[gp path] (4.657,4.771)--(4.477,4.771);
\node[gp node right] at (0.741,4.771) {$10^{0}$};
\draw[gp path] (0.855,0.729)--(0.855,0.909);
\draw[gp path] (0.855,4.771)--(0.855,4.591);
\node[gp node center] at (0.855,0.501) {$10^{0}$};
\draw[gp path] (1.147,0.729)--(1.147,0.819);
\draw[gp path] (1.147,4.771)--(1.147,4.681);
\draw[gp path] (1.440,0.729)--(1.440,0.909);
\draw[gp path] (1.440,4.771)--(1.440,4.591);
\node[gp node center] at (1.440,0.501) {$10^{2}$};
\draw[gp path] (1.732,0.729)--(1.732,0.819);
\draw[gp path] (1.732,4.771)--(1.732,4.681);
\draw[gp path] (2.025,0.729)--(2.025,0.909);
\draw[gp path] (2.025,4.771)--(2.025,4.591);
\node[gp node center] at (2.025,0.501) {$10^{4}$};
\draw[gp path] (2.317,0.729)--(2.317,0.819);
\draw[gp path] (2.317,4.771)--(2.317,4.681);
\draw[gp path] (2.610,0.729)--(2.610,0.909);
\draw[gp path] (2.610,4.771)--(2.610,4.591);
\node[gp node center] at (2.610,0.501) {$10^{6}$};
\draw[gp path] (2.902,0.729)--(2.902,0.819);
\draw[gp path] (2.902,4.771)--(2.902,4.681);
\draw[gp path] (3.195,0.729)--(3.195,0.909);
\draw[gp path] (3.195,4.771)--(3.195,4.591);
\node[gp node center] at (3.195,0.501) {$10^{8}$};
\draw[gp path] (3.487,0.729)--(3.487,0.819);
\draw[gp path] (3.487,4.771)--(3.487,4.681);
\draw[gp path] (3.780,0.729)--(3.780,0.909);
\draw[gp path] (3.780,4.771)--(3.780,4.591);
\node[gp node center] at (3.780,0.501) {$10^{10}$};
\draw[gp path] (4.072,0.729)--(4.072,0.819);
\draw[gp path] (4.072,4.771)--(4.072,4.681);
\draw[gp path] (4.365,0.729)--(4.365,0.909);
\draw[gp path] (4.365,4.771)--(4.365,4.591);
\node[gp node center] at (4.365,0.501) {$10^{12}$};
\draw[gp path] (4.657,0.729)--(4.657,0.819);
\draw[gp path] (4.657,4.771)--(4.657,4.681);
\draw[gp path] (0.855,4.771)--(0.855,0.729)--(4.657,0.729)--(4.657,4.771)--cycle;
\gpcolor{rgb color={0.000,0.620,0.451}}
\node[gp node left] at (3.173,4.452) {breakdown!};
\gpcolor{rgb color={0.337,0.706,0.914}}
\node[gp node right] at (3.601,2.529) {different $\kappa(X,W)$};
\gpcolor{color=gp lt color border}
\node[gp node center] at (2.756,0.159) {condition number $\kappa(X)$};
\gpcolor{rgb color={0.580,0.000,0.827}}
\gpsetlinewidth{2.00}
\draw[gp path] (0.855,1.513)--(1.732,1.490)--(2.610,1.532)--(3.487,1.506)--(4.365,1.532);
\draw[gp path] (0.855,1.485)--(1.732,1.550)--(2.610,1.517)--(3.487,1.506)--(4.365,1.480);
\draw[gp path] (0.855,1.498)--(1.732,1.510)--(2.610,1.520)--(3.487,1.506)--(4.365,1.513);
\gpsetpointsize{8.00}
\gppoint{gp mark 1}{(0.855,1.513)}
\gppoint{gp mark 1}{(1.732,1.490)}
\gppoint{gp mark 1}{(2.610,1.532)}
\gppoint{gp mark 1}{(3.487,1.506)}
\gppoint{gp mark 1}{(4.365,1.532)}
\gppoint{gp mark 1}{(0.855,1.485)}
\gppoint{gp mark 1}{(1.732,1.550)}
\gppoint{gp mark 1}{(2.610,1.517)}
\gppoint{gp mark 1}{(3.487,1.506)}
\gppoint{gp mark 1}{(4.365,1.480)}
\gppoint{gp mark 1}{(0.855,1.498)}
\gppoint{gp mark 1}{(1.732,1.510)}
\gppoint{gp mark 1}{(2.610,1.520)}
\gppoint{gp mark 1}{(3.487,1.506)}
\gppoint{gp mark 1}{(4.365,1.513)}
\gpcolor{rgb color={0.000,0.620,0.451}}
\draw[gp path] (0.855,1.676)--(1.732,2.922)--(2.610,4.309)--(3.487,4.771)--(4.365,4.771);
\draw[gp path] (0.855,1.633)--(1.732,2.939)--(2.610,4.264)--(3.487,4.771)--(4.365,4.771);
\draw[gp path] (0.855,1.603)--(1.732,3.003)--(2.610,4.306)--(3.487,4.711)--(4.365,4.704);
\gppoint{gp mark 2}{(0.855,1.676)}
\gppoint{gp mark 2}{(1.732,2.922)}
\gppoint{gp mark 2}{(2.610,4.309)}
\gppoint{gp mark 2}{(3.487,4.771)}
\gppoint{gp mark 2}{(4.365,4.771)}
\gppoint{gp mark 2}{(0.855,1.633)}
\gppoint{gp mark 2}{(1.732,2.939)}
\gppoint{gp mark 2}{(2.610,4.264)}
\gppoint{gp mark 2}{(3.487,4.771)}
\gppoint{gp mark 2}{(4.365,4.771)}
\gppoint{gp mark 2}{(0.855,1.603)}
\gppoint{gp mark 2}{(1.732,3.003)}
\gppoint{gp mark 2}{(2.610,4.306)}
\gppoint{gp mark 2}{(3.487,4.711)}
\gppoint{gp mark 2}{(4.365,4.704)}
\gpcolor{rgb color={0.337,0.706,0.914}}
\draw[gp path] (0.855,0.879)--(1.732,0.871)--(2.610,0.910)--(3.487,2.218)--(4.365,3.654);
\draw[gp path] (0.855,0.889)--(1.732,0.898)--(2.610,0.908)--(3.487,2.226)--(4.365,3.051);
\draw[gp path] (0.855,0.882)--(1.732,0.906)--(2.610,0.899)--(3.487,1.677)--(4.365,1.674);
\gppoint{gp mark 3}{(0.855,0.879)}
\gppoint{gp mark 3}{(1.732,0.871)}
\gppoint{gp mark 3}{(2.610,0.910)}
\gppoint{gp mark 3}{(3.487,2.218)}
\gppoint{gp mark 3}{(4.365,3.654)}
\gppoint{gp mark 3}{(0.855,0.889)}
\gppoint{gp mark 3}{(1.732,0.898)}
\gppoint{gp mark 3}{(2.610,0.908)}
\gppoint{gp mark 3}{(3.487,2.226)}
\gppoint{gp mark 3}{(4.365,3.051)}
\gppoint{gp mark 3}{(0.855,0.882)}
\gppoint{gp mark 3}{(1.732,0.906)}
\gppoint{gp mark 3}{(2.610,0.899)}
\gppoint{gp mark 3}{(3.487,1.677)}
\gppoint{gp mark 3}{(4.365,1.674)}
\gpcolor{color=gp lt color border}
\gpsetlinewidth{1.00}
\draw[gp path] (0.855,4.771)--(0.855,0.729)--(4.657,0.729)--(4.657,4.771)--cycle;
\gpdefrectangularnode{gp plot 1}{\pgfpoint{0.855cm}{0.729cm}}{\pgfpoint{4.657cm}{4.771cm}}
\end{tikzpicture}
    \end{tabular}
  \end{center}
  \caption{Accuracy after one iteration (phase~I and phase~II) for
           synthetic test matrices $W \in \R^{n \times k}$,
           $X \in \R^{n \times n_{b}}$, where
           $n = 1000$, $k = 20$, and $n_{b} = 4$.}
  \label{fig:JDAccuracyDD}
\end{figure}

  Figure~\ref{fig:JDAccuracyDD}
shows the results of a single two-phase orthogonalization, without
iteration, for synthetic test matrices with varying condition.
If $X$ is ill-conditioned then TSQR does a much better job on $X$ than
SVQB, but this does not carry over to orthogonality against $W$,
and using DD kernels can improve both orthogonalities by at least two
orders of magnitude.

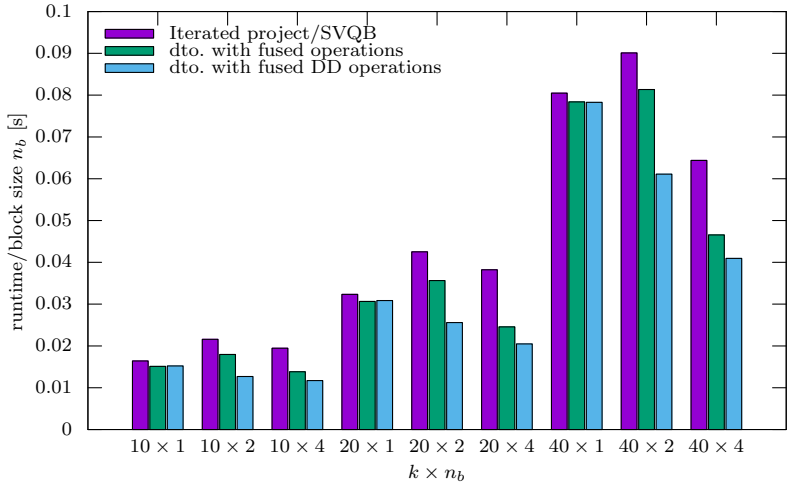
\begin{figure}
  \centerline{%
    \begin{tikzpicture}[gnuplot]
\tikzset{every node/.append style={scale=0.9}}
\path (0.000,0.000) rectangle (10.500,6.500);
\gpcolor{color=gp lt color border}
\gpsetlinetype{gp lt border}
\gpsetlinewidth{1.00}
\draw[gp path] (0.969,0.729)--(1.149,0.729);
\draw[gp path] (10.157,0.729)--(9.977,0.729);
\node[gp node right] at (0.855,0.729) {$0$};
\draw[gp path] (0.969,1.283)--(1.149,1.283);
\draw[gp path] (10.157,1.283)--(9.977,1.283);
\node[gp node right] at (0.855,1.283) {$0.01$};
\draw[gp path] (0.969,1.837)--(1.149,1.837);
\draw[gp path] (10.157,1.837)--(9.977,1.837);
\node[gp node right] at (0.855,1.837) {$0.02$};
\draw[gp path] (0.969,2.392)--(1.149,2.392);
\draw[gp path] (10.157,2.392)--(9.977,2.392);
\node[gp node right] at (0.855,2.392) {$0.03$};
\draw[gp path] (0.969,2.946)--(1.149,2.946);
\draw[gp path] (10.157,2.946)--(9.977,2.946);
\node[gp node right] at (0.855,2.946) {$0.04$};
\draw[gp path] (0.969,3.500)--(1.149,3.500);
\draw[gp path] (10.157,3.500)--(9.977,3.500);
\node[gp node right] at (0.855,3.500) {$0.05$};
\draw[gp path] (0.969,4.054)--(1.149,4.054);
\draw[gp path] (10.157,4.054)--(9.977,4.054);
\node[gp node right] at (0.855,4.054) {$0.06$};
\draw[gp path] (0.969,4.608)--(1.149,4.608);
\draw[gp path] (10.157,4.608)--(9.977,4.608);
\node[gp node right] at (0.855,4.608) {$0.07$};
\draw[gp path] (0.969,5.163)--(1.149,5.163);
\draw[gp path] (10.157,5.163)--(9.977,5.163);
\node[gp node right] at (0.855,5.163) {$0.08$};
\draw[gp path] (0.969,5.717)--(1.149,5.717);
\draw[gp path] (10.157,5.717)--(9.977,5.717);
\node[gp node right] at (0.855,5.717) {$0.09$};
\draw[gp path] (0.969,6.271)--(1.149,6.271);
\draw[gp path] (10.157,6.271)--(9.977,6.271);
\node[gp node right] at (0.855,6.271) {$0.1$};
\draw[gp path] (1.888,0.729)--(1.888,0.909);
\draw[gp path] (1.888,6.271)--(1.888,6.091);
\node[gp node center] at (1.888,0.501) {$10\times1$};
\draw[gp path] (2.807,0.729)--(2.807,0.909);
\draw[gp path] (2.807,6.271)--(2.807,6.091);
\node[gp node center] at (2.807,0.501) {$10\times2$};
\draw[gp path] (3.725,0.729)--(3.725,0.909);
\draw[gp path] (3.725,6.271)--(3.725,6.091);
\node[gp node center] at (3.725,0.501) {$10\times4$};
\draw[gp path] (4.644,0.729)--(4.644,0.909);
\draw[gp path] (4.644,6.271)--(4.644,6.091);
\node[gp node center] at (4.644,0.501) {$20\times1$};
\draw[gp path] (5.563,0.729)--(5.563,0.909);
\draw[gp path] (5.563,6.271)--(5.563,6.091);
\node[gp node center] at (5.563,0.501) {$20\times2$};
\draw[gp path] (6.482,0.729)--(6.482,0.909);
\draw[gp path] (6.482,6.271)--(6.482,6.091);
\node[gp node center] at (6.482,0.501) {$20\times4$};
\draw[gp path] (7.401,0.729)--(7.401,0.909);
\draw[gp path] (7.401,6.271)--(7.401,6.091);
\node[gp node center] at (7.401,0.501) {$40\times1$};
\draw[gp path] (8.319,0.729)--(8.319,0.909);
\draw[gp path] (8.319,6.271)--(8.319,6.091);
\node[gp node center] at (8.319,0.501) {$40\times2$};
\draw[gp path] (9.238,0.729)--(9.238,0.909);
\draw[gp path] (9.238,6.271)--(9.238,6.091);
\node[gp node center] at (9.238,0.501) {$40\times4$};
\draw[gp path] (0.969,6.271)--(0.969,0.729)--(10.157,0.729)--(10.157,6.271)--cycle;
\node[gp node center,rotate=-270] at (0.051,3.500) {runtime/block size $n_b$ [s]};
\node[gp node center] at (5.563,0.159) {$k \times n_b$};
\node[gp node left] at (1.947,5.977) {Iterated project/SVQB};
\gpfill{rgb color={0.580,0.000,0.827}} (1.197,5.920)--(1.833,5.920)--(1.833,6.034)--(1.197,6.034)--cycle;
\draw[gp path] (1.197,5.920)--(1.833,5.920)--(1.833,6.034)--(1.197,6.034)--cycle;
\gpfill{rgb color={0.580,0.000,0.827}} (1.555,0.729)--(1.762,0.729)--(1.762,1.639)--(1.555,1.639)--cycle;
\draw[gp path] (1.555,0.729)--(1.555,1.638)--(1.761,1.638)--(1.761,0.729)--cycle;
\gpfill{rgb color={0.580,0.000,0.827}} (2.474,0.729)--(2.681,0.729)--(2.681,1.926)--(2.474,1.926)--cycle;
\draw[gp path] (2.474,0.729)--(2.474,1.925)--(2.680,1.925)--(2.680,0.729)--cycle;
\gpfill{rgb color={0.580,0.000,0.827}} (3.392,0.729)--(3.600,0.729)--(3.600,1.808)--(3.392,1.808)--cycle;
\draw[gp path] (3.392,0.729)--(3.392,1.807)--(3.599,1.807)--(3.599,0.729)--cycle;
\gpfill{rgb color={0.580,0.000,0.827}} (4.311,0.729)--(4.519,0.729)--(4.519,2.522)--(4.311,2.522)--cycle;
\draw[gp path] (4.311,0.729)--(4.311,2.521)--(4.518,2.521)--(4.518,0.729)--cycle;
\gpfill{rgb color={0.580,0.000,0.827}} (5.230,0.729)--(5.438,0.729)--(5.438,3.086)--(5.230,3.086)--cycle;
\draw[gp path] (5.230,0.729)--(5.230,3.085)--(5.437,3.085)--(5.437,0.729)--cycle;
\gpfill{rgb color={0.580,0.000,0.827}} (6.149,0.729)--(6.356,0.729)--(6.356,2.848)--(6.149,2.848)--cycle;
\draw[gp path] (6.149,0.729)--(6.149,2.847)--(6.355,2.847)--(6.355,0.729)--cycle;
\gpfill{rgb color={0.580,0.000,0.827}} (7.068,0.729)--(7.275,0.729)--(7.275,5.190)--(7.068,5.190)--cycle;
\draw[gp path] (7.068,0.729)--(7.068,5.189)--(7.274,5.189)--(7.274,0.729)--cycle;
\gpfill{rgb color={0.580,0.000,0.827}} (7.986,0.729)--(8.194,0.729)--(8.194,5.723)--(7.986,5.723)--cycle;
\draw[gp path] (7.986,0.729)--(7.986,5.722)--(8.193,5.722)--(8.193,0.729)--cycle;
\gpfill{rgb color={0.580,0.000,0.827}} (8.905,0.729)--(9.113,0.729)--(9.113,4.298)--(8.905,4.298)--cycle;
\draw[gp path] (8.905,0.729)--(8.905,4.297)--(9.112,4.297)--(9.112,0.729)--cycle;
\node[gp node left] at (1.947,5.749) {dto.\ with fused operations};
\gpfill{rgb color={0.000,0.620,0.451}} (1.197,5.692)--(1.833,5.692)--(1.833,5.806)--(1.197,5.806)--cycle;
\draw[gp path] (1.197,5.692)--(1.833,5.692)--(1.833,5.806)--(1.197,5.806)--cycle;
\gpfill{rgb color={0.000,0.620,0.451}} (1.784,0.729)--(1.992,0.729)--(1.992,1.567)--(1.784,1.567)--cycle;
\draw[gp path] (1.784,0.729)--(1.784,1.566)--(1.991,1.566)--(1.991,0.729)--cycle;
\gpfill{rgb color={0.000,0.620,0.451}} (2.703,0.729)--(2.911,0.729)--(2.911,1.725)--(2.703,1.725)--cycle;
\draw[gp path] (2.703,0.729)--(2.703,1.724)--(2.910,1.724)--(2.910,0.729)--cycle;
\gpfill{rgb color={0.000,0.620,0.451}} (3.622,0.729)--(3.830,0.729)--(3.830,1.496)--(3.622,1.496)--cycle;
\draw[gp path] (3.622,0.729)--(3.622,1.495)--(3.829,1.495)--(3.829,0.729)--cycle;
\gpfill{rgb color={0.000,0.620,0.451}} (4.541,0.729)--(4.749,0.729)--(4.749,2.428)--(4.541,2.428)--cycle;
\draw[gp path] (4.541,0.729)--(4.541,2.427)--(4.748,2.427)--(4.748,0.729)--cycle;
\gpfill{rgb color={0.000,0.620,0.451}} (5.460,0.729)--(5.667,0.729)--(5.667,2.705)--(5.460,2.705)--cycle;
\draw[gp path] (5.460,0.729)--(5.460,2.704)--(5.666,2.704)--(5.666,0.729)--cycle;
\gpfill{rgb color={0.000,0.620,0.451}} (6.378,0.729)--(6.586,0.729)--(6.586,2.091)--(6.378,2.091)--cycle;
\draw[gp path] (6.378,0.729)--(6.378,2.090)--(6.585,2.090)--(6.585,0.729)--cycle;
\gpfill{rgb color={0.000,0.620,0.451}} (7.297,0.729)--(7.505,0.729)--(7.505,5.074)--(7.297,5.074)--cycle;
\draw[gp path] (7.297,0.729)--(7.297,5.073)--(7.504,5.073)--(7.504,0.729)--cycle;
\gpfill{rgb color={0.000,0.620,0.451}} (8.216,0.729)--(8.424,0.729)--(8.424,5.238)--(8.216,5.238)--cycle;
\draw[gp path] (8.216,0.729)--(8.216,5.237)--(8.423,5.237)--(8.423,0.729)--cycle;
\gpfill{rgb color={0.000,0.620,0.451}} (9.135,0.729)--(9.343,0.729)--(9.343,3.310)--(9.135,3.310)--cycle;
\draw[gp path] (9.135,0.729)--(9.135,3.309)--(9.342,3.309)--(9.342,0.729)--cycle;
\node[gp node left] at (1.947,5.521) {dto.\ with fused DD operations};
\gpfill{rgb color={0.337,0.706,0.914}} (1.197,5.464)--(1.833,5.464)--(1.833,5.578)--(1.197,5.578)--cycle;
\draw[gp path] (1.197,5.464)--(1.833,5.464)--(1.833,5.578)--(1.197,5.578)--cycle;
\gpfill{rgb color={0.337,0.706,0.914}} (2.014,0.729)--(2.222,0.729)--(2.222,1.573)--(2.014,1.573)--cycle;
\draw[gp path] (2.014,0.729)--(2.014,1.572)--(2.221,1.572)--(2.221,0.729)--cycle;
\gpfill{rgb color={0.337,0.706,0.914}} (2.933,0.729)--(3.141,0.729)--(3.141,1.433)--(2.933,1.433)--cycle;
\draw[gp path] (2.933,0.729)--(2.933,1.432)--(3.140,1.432)--(3.140,0.729)--cycle;
\gpfill{rgb color={0.337,0.706,0.914}} (3.852,0.729)--(4.059,0.729)--(4.059,1.379)--(3.852,1.379)--cycle;
\draw[gp path] (3.852,0.729)--(3.852,1.378)--(4.058,1.378)--(4.058,0.729)--cycle;
\gpfill{rgb color={0.337,0.706,0.914}} (4.771,0.729)--(4.978,0.729)--(4.978,2.440)--(4.771,2.440)--cycle;
\draw[gp path] (4.771,0.729)--(4.771,2.439)--(4.977,2.439)--(4.977,0.729)--cycle;
\gpfill{rgb color={0.337,0.706,0.914}} (5.689,0.729)--(5.897,0.729)--(5.897,2.148)--(5.689,2.148)--cycle;
\draw[gp path] (5.689,0.729)--(5.689,2.147)--(5.896,2.147)--(5.896,0.729)--cycle;
\gpfill{rgb color={0.337,0.706,0.914}} (6.608,0.729)--(6.816,0.729)--(6.816,1.865)--(6.608,1.865)--cycle;
\draw[gp path] (6.608,0.729)--(6.608,1.864)--(6.815,1.864)--(6.815,0.729)--cycle;
\gpfill{rgb color={0.337,0.706,0.914}} (7.527,0.729)--(7.735,0.729)--(7.735,5.068)--(7.527,5.068)--cycle;
\draw[gp path] (7.527,0.729)--(7.527,5.067)--(7.734,5.067)--(7.734,0.729)--cycle;
\gpfill{rgb color={0.337,0.706,0.914}} (8.446,0.729)--(8.653,0.729)--(8.653,4.116)--(8.446,4.116)--cycle;
\draw[gp path] (8.446,0.729)--(8.446,4.115)--(8.652,4.115)--(8.652,0.729)--cycle;
\gpfill{rgb color={0.337,0.706,0.914}} (9.365,0.729)--(9.572,0.729)--(9.572,2.999)--(9.365,2.999)--cycle;
\draw[gp path] (9.365,0.729)--(9.365,2.998)--(9.571,2.998)--(9.571,0.729)--cycle;
\draw[gp path] (0.969,6.271)--(0.969,0.729)--(10.157,0.729)--(10.157,6.271)--cycle;
\gpdefrectangularnode{gp plot 1}{\pgfpoint{0.969cm}{0.729cm}}{\pgfpoint{10.157cm}{6.271cm}}
\end{tikzpicture}
}
  \caption{Runtime ``per vector'' for overall orthogonalization
           of $X \in \R^{n \times n_{b}}$ against
           $W \in \R^{n \times k}$ and $X$
           (phases~I and II, iterated until convergence at
           $\epsilon=10^{-10}$) on an Intel Haswell workstation.
           $\kappa( X ) = 10^{-6}$,
           $\kappa( X, W ) = 10^{-12}$,
           $n = 8 \cdot 10^{6}$,
           $k$ and $n_{b}$ are indicated on the horizontal axis.}
  \label{fig:JDRuntime}
\end{figure}

On modern architectures, even the performance of matrix--matrix
multiplications such as $\tilde V^{T} \tilde V$ is memory-bound if the
matrix $\tilde V \in \R^{n \times n_{b}}$ is only very few columns wide.
Then the additional arithmetic operations required in the DD kernels
come almost for free, and operations on small $n_{b} \times n_{b}$
matrices are cost negligible, even in extended precision.
  Figure~\ref{fig:JDRuntime}
compares timings for the overall orthogonalization with a
straight-forward implementation, one that uses kernel fusion (combining
several basic operations to further reduce memory accesses; not
discussed here), and one with fused DD kernels.
It reveals that using DD routines can even reduce overall time because
the higher accuracy achieved in each iteration can lead to a lower
number of iterations to reach convergence.

This technique can be useful for any algorithm that requires
orthogonalizing a set $X$ of vectors with respect to themselves and to
another set $W$ of (already orthonormal) vectors.
It also extends to $B$-inner products, which is important, e.g., when
solving generalized eigenvalue problems.


\subsection{Mixed precision in SCF cycles with ELPA-AEO}%
  \label{sec:MixedPrecisionInELPAAEO}

\newcommand{\dr}{\mathrm{d}\vec{r}\,}

The solution of the quantum-mechanical electronic-structure problem is
at the basis of studies in computational chemistry, solid state physics,
and materials science. 
In density-functional theory (DFT), the most wide-spread
electronic-structure formalism, this implies finding the electronic
density $n(\vec{r})$ that minimizes ($E_0=\min E[n(\vec{r})]$) the
convex total-energy functional $E[n(\vec{r})]$ under the constraint that
the number of electrons, $N=\int \dr n(\vec{r})$, is conserved.
Here, the set of $3 M$ nuclear coordinates $\{\vec{R}\}$ enters
$E[n(\vec{r})]$ parametrically. 
Formally, this variational problem requires to find the stationary
solution of the eigenvalue problem (EVP)
\begin{equation}
  H[n(\vec{r})] \, \Psi(\vec{r}) = \varepsilon \Psi(\vec{r})
  \quad \text{ with } \quad
  n(\vec{r}) = \sum_{s=1}^N  |\Psi_s(\vec{r})|^2
  \label{eq:SG}
\end{equation}
in Hilbert space by iteratively updating $n(\vec{r})$, which depends on
the $N$~eigenstates $\Psi_s$ with the lowest eigenvalues
$\varepsilon_s$.
This so called self-consistent field (SCF) cycle runs until
``self-consistency'' is achieved, i.e., until the mean interaction field
contained in $H[n_k(\vec{r})]$ and/or other quantities (see below) do
not change substantially between iterations anymore.
In each step of the SCF cycle, the integro-differential equation
  \eqref{eq:SG}
has to be solved. In practice, this is done by algebraizing
  Eq.~\eqref{eq:SG}  
via a basis set expansion $\Psi_s=\sum_i x_{si} \varphi_i(\vec{r})$ of
the so called orbitals in terms of appropriately chosen basis functions
$\varphi_i(\vec{r})$, e.g., plane waves, localized functions, etc. 
By this means, one obtains a generalized EVP
\[
  A[n(\vec{r})] \, x = \lambda B x \; ,
\]
in which the Hamiltonian $A$ and the overlap matrix $B$ are defined as
\begin{equation}
  A_{ij}[n(\vec{r})]
  =
  \int \dr \varphi_i^*(\vec{r}) \, H[n(\vec{r})] \, \varphi_j(\vec{r})
  \quad \text{ and } \quad
  B_{ij} = \int \dr \varphi_i^*(\vec{r}) \, \varphi_j(\vec{r}) \; .
  \label{eq:Def}
\end{equation}
As becomes clear from
  Eq.~\eqref{eq:Def},
the size of the EVP is thus determined by the number $K$ of basis
functions $\varphi_i(\vec{r})$ employed in the calculation.
For efficient, atom-centered basis functions the ratio $N/K$ of required
eigenstates to matrix dimension typically ranges between $10$ and
$50$\%, rendering a direct solver competitive.

One SCF cycle yields the total energy $E_0(\{\vec{R}\})$ for just one
set of nuclear coordinates $\{\vec{R}\}$.
Studying molecules and materials requires the exploration of the high
dimensional potential-energy surface (PES) which is given by
$E_0(\{\vec{R}\})$ as a function of $\{\vec{R}\}$, e.g.,
via molecular dynamics (MD), statistical (e.g. Monte Carlo) sampling,
or minimization and saddle point search algorithms. 
Accordingly, a typical computational study requires thousands if not
millions of SCF cycles (about $10$--$100$ SCF steps per cycle) to be
performed in a single simulation.
This large number of SCF steps makes it mandatory to investigate
strategies to reduce the computational effort. 
Since only the final result of each converged SCF cycle is of physical
relevance at all, the SCF procedure can be accelerated by using single
precision (SP) routines instead of double precision (DP) ones in the
appropriate eigensolver steps (cf.\
  Section~\ref{sec:ELPA-AEO}),
as long as the final converged result is not altered up to the precision
mandated by the problem at hand.
The eigensolver steps discussed in this section are the Cholesky
decomposition (i), the transformation to the standard eigenproblem (ii),
and its standard diagonalization, which combines
tridiagonalization~(iii) and the tridiagonal eigensolver (iv), as
defined in
  Section~\ref{sec:ELPA-AEO}.

\begin{figure}
  \centerline{%
    \includegraphics[width=0.5\textwidth]{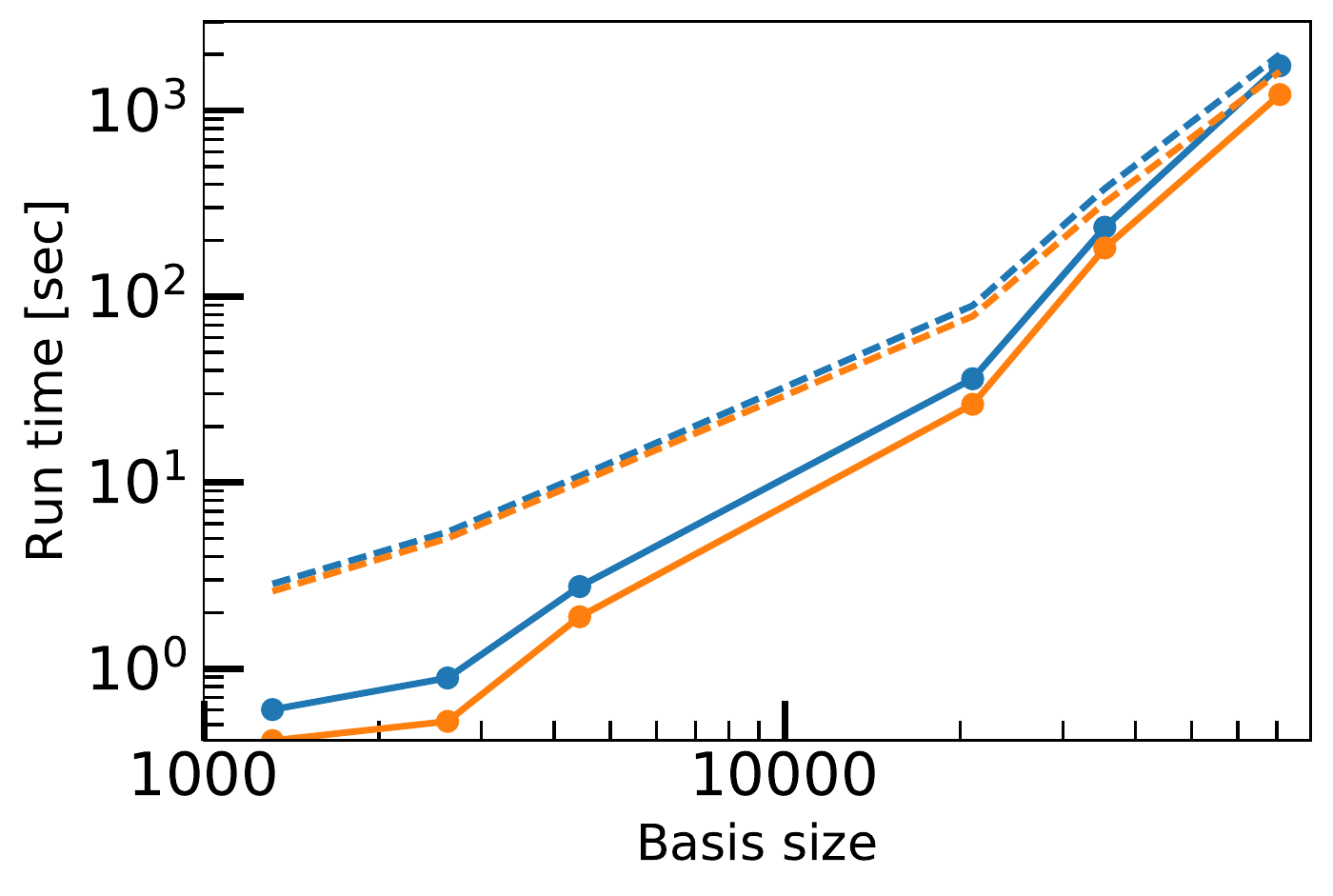}}
  \caption{Total time for one SCF cycle (dashed lines) and total time
    spent in ELPA-AEO (solid lines) as function of the basis set size
    using SP (orange) and DP routines (blue) for zirconia (ZrO$_2$,
    $M$ from $6$ to $768$ atoms, $N$ from $112$ to $14,\!336$
    electrons). 
    The calculations were performed with $8$ Intel Xeon E5-2698v3 CPUs
    ($4$ nodes, $32$ cores/node) and $4$~GB of RAM per core.}
  \label{fig:FHI1}
\end{figure}

\begin{figure}
  \centerline{%
    \includegraphics[width=0.45\textwidth]{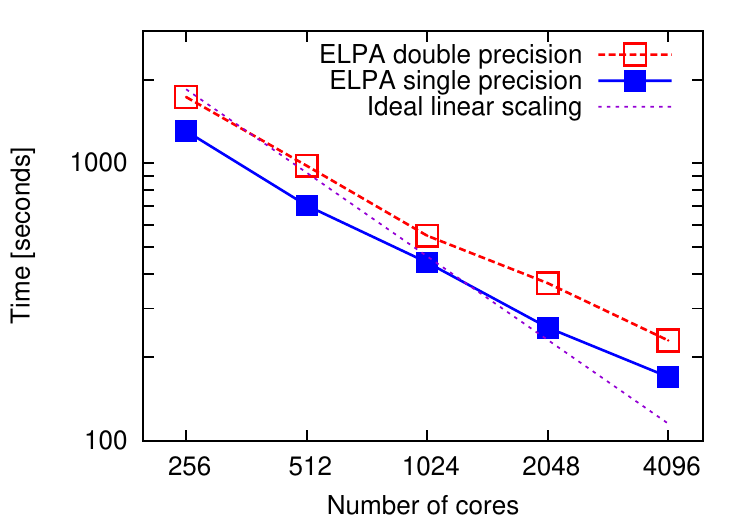}
    \qquad
    \raisebox{1.5em}{%
      \includegraphics[width=0.45\textwidth]{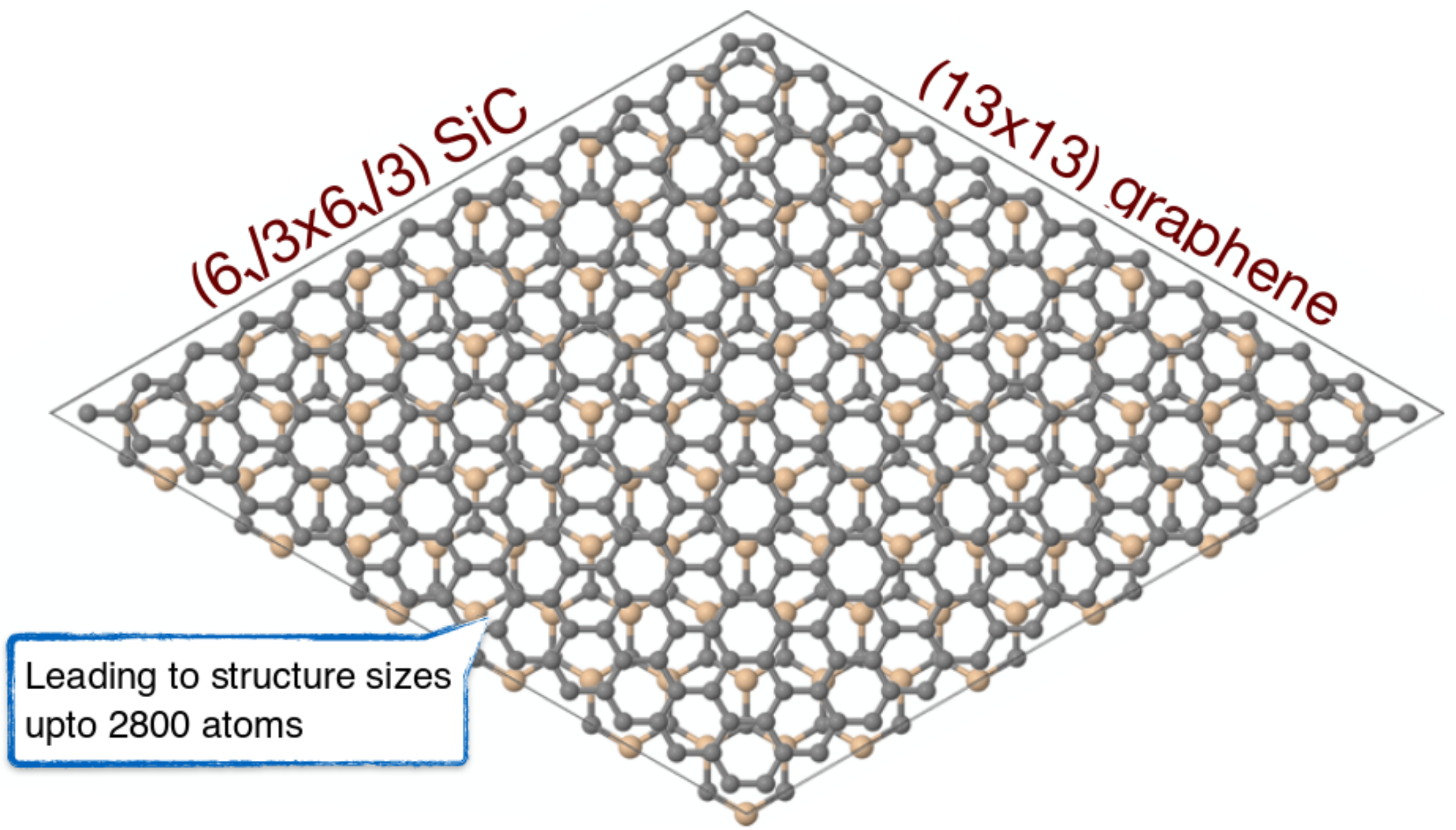}}}
  \caption{Time for solving the EVP for Zero-Layer Graphene
    ($M=1,\!742$, $65,\!346$ basis functions, LDA, $\Gamma$-point only)
    as function of the number of cores with DP (empty squares) and
    SP (filled squares).
    The calculations were performed with the IBM iDataPlex HPC system
    HYDRA using $2.6$~GHz Intel Sandy Bridge EP nodes with $16$ cores
    per node.}
  \label{fig:7}
\end{figure}

To showcase the importance of the readily available SP routines in
ELPA-AEO, we have performed DFT calculations with the all-electron,
numeric atomic orbitals based code FHI-aims~%
  \cite{Blum:2009fe},
which supports both ELPA and ELPA-AEO through the ELSI package~%
  \cite{Yu:2018ih}.
For this purpose, we have run benchmark calculations for zirconia
(ZrO$_2$) in its tetragonal polymorph, a wide band-gap insulator often
employed as thermal insulator in aeronautic applications~%
  \cite{Carbogno:2014wa,Carbogno:2017gc}.
Supercells containing between $M = 6$ and $768$~atoms ($N = 112$ and
$14,\!336$ electrons) were investigated using the PBEsol
exchange-correlation functional, ``light'' defaults for the numerical
settings, and chemical species-specific ``Tier 1'' defaults for the
basis functions $\varphi_{i}$.
Accordingly, this translates to  basis sets yielding matrix dimensions
from $K = 1,\!312$ to $70,\!848$ for the investigated systems.
The finite $\vec{k}$-point grid required to sample reciprocal space to
model such extended materials using periodic boundary conditions was
chosen in such a way that the $\vec{k}$-point density is roughly
constant (between $128$ and $216$ $\vec{k}$-points in the respective
primitive Brillouin zone). 
As an example,
  Figure~\ref{fig:FHI1}
shows the total time for one SCF step and the total time spent in
solving the EVP with SP and DP as function of the system size.
Here, SP is only used in the diagonalization (steps~(iii) and (iv)
introduced in
  Section~\ref{sec:ELPA-AEO}).
For larger system sizes (more than $10^4$ basis functions), the
computational time spent in the calculation of $A[n(\vec{r})]$, which
typically exhibits linear scaling with respect to $N$ in FHI-aims~%
  \cite{Havu:2009ug},
becomes increasingly negligible compared to the EVP, which starts to
dominate the computational time due to its cubic scaling.
Switching from DP to SP thus allows for computational savings in the
solution of the EVP on the order of $30$--$50$\%.
Even for medium system sizes ($M = 96$ with $K = 2,\!624$ basis
functions) that are routinely addressed in DFT calculations~%
  \cite{Carbogno:2017gc}
this already translates into savings in total computational time of
around $10$\%, while savings of more than $20$\% are observed for larger
systems (up to over $40\%$ in
  Figure~\ref{fig:7}).

\begin{figure}
  \centerline{%
    \includegraphics[width=0.45\textwidth]%
                    {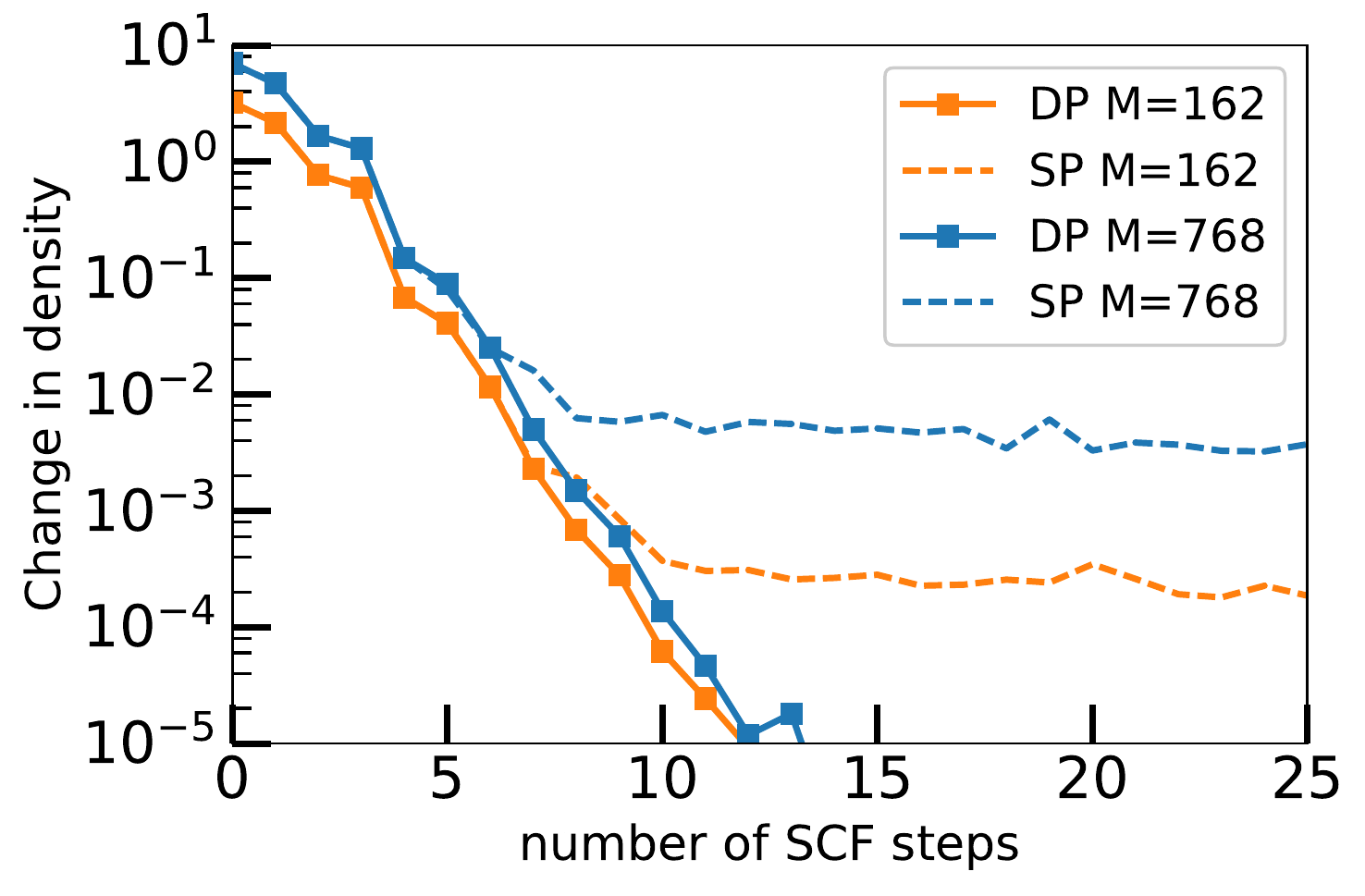}
    \qquad
    \includegraphics[width=0.45\textwidth]%
                    {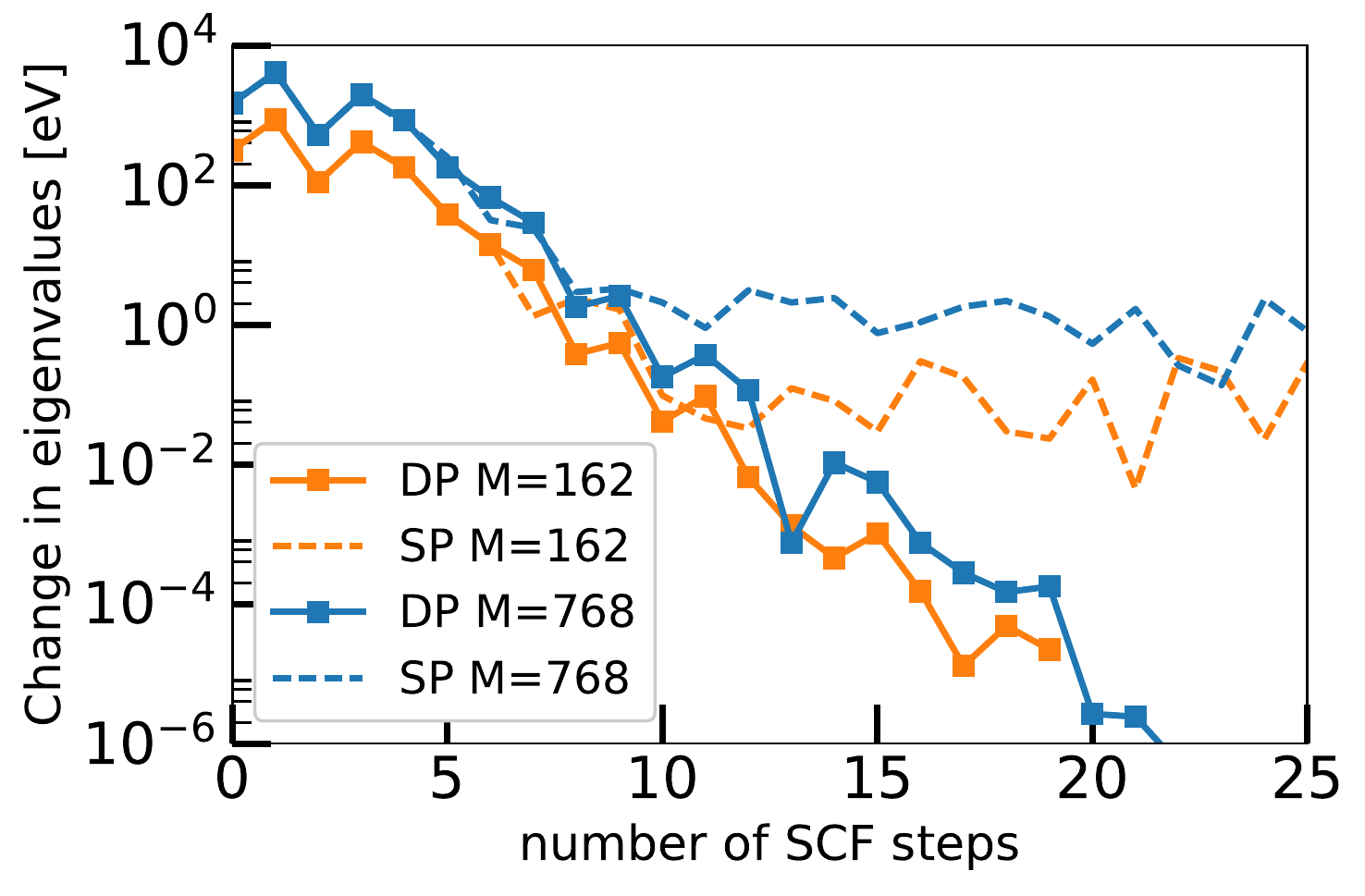}}
  \caption{Change in density (left) and eigenvalues (right) as function
    of the number of SCF steps during a full SCF cycle when
    single-(squares) or double-precision~(triangles) routines are used.
    Calculations were performed for zirconia ($M=162$ and $M=768$ atoms,
    $N = 3,\!024$ and $14,\!336$ electrons,
    respectively, using the settings of
      Figure~\ref{fig:FHI1}).}
  \label{fig:FHI2}  
\end{figure}

However, SP routines cannot be exploited during the full SCF cycle:
once a certain accuracy is reached, further SP SCF iterations do no
longer approach convergence.
This is demonstrated for ZrO$_2$ in
  Figure~\ref{fig:FHI2}. 
In each SCF step, we monitor two properties that are typically used for
determining the convergence of such calculations:
(I) the change in charge density between two subsequent steps $k$ and
  $k+1$, $\Delta n = \int \dr |n_k(\vec{r})-n_{k+1}(\vec{r})|$, and
(II) the change in the sum of the $N$ lowest eigenvalues,
 $\Delta \varepsilon
  =
  \sum_{s=1}^N\varepsilon_s^{(k)} - \varepsilon_s^{(k+1)}$.
For $M=162$ atoms, we observe that $\Delta n$ stalls at approximately
$2.5\cdot10^{-4}$ electrons after the $10$th SCF iteration for the
calculation using SP.
Similarly, $\Delta \varepsilon$ stalls at a value of $5\cdot10^{-2}$~eV,
showing a less regular behavior, both in SP and DP.
This can be traced back to the fact that the total-energy functional is
not variational with respect to the eigenvalues.
As also shown in
  Figure~\ref{fig:FHI2}
for $M=768$ atoms ($N = 14,\!336$ electrons), the observed thresholds at
which using SP no longer guarantees approaching convergence is, however,
system and  size dependent, since the respective quantities (energy,
density, sum of eigenvalues, etc.) are extensive with system size, i.e.,
they scale linearly with the number of electrons, $N$.
For these reasons, convergence criteria in DFT calculations are
typically not chosen with respect to extensive quantities as the total
energy, but with respect to intensive quantities, such as the total
energy per atom.
Hence the fraction of iterations for which SP routines can be used
($>30$\%) are roughly independent of the system size, given that both
the target quantity and its change, e.g., $n(\vec{r})$ and
$\Delta n(\vec{r})$, are extensive with system size. 

\begin{figure}
  \centerline{%
    \begin{tabular}{cc}
        \includegraphics[width=0.45\textwidth]%
                        {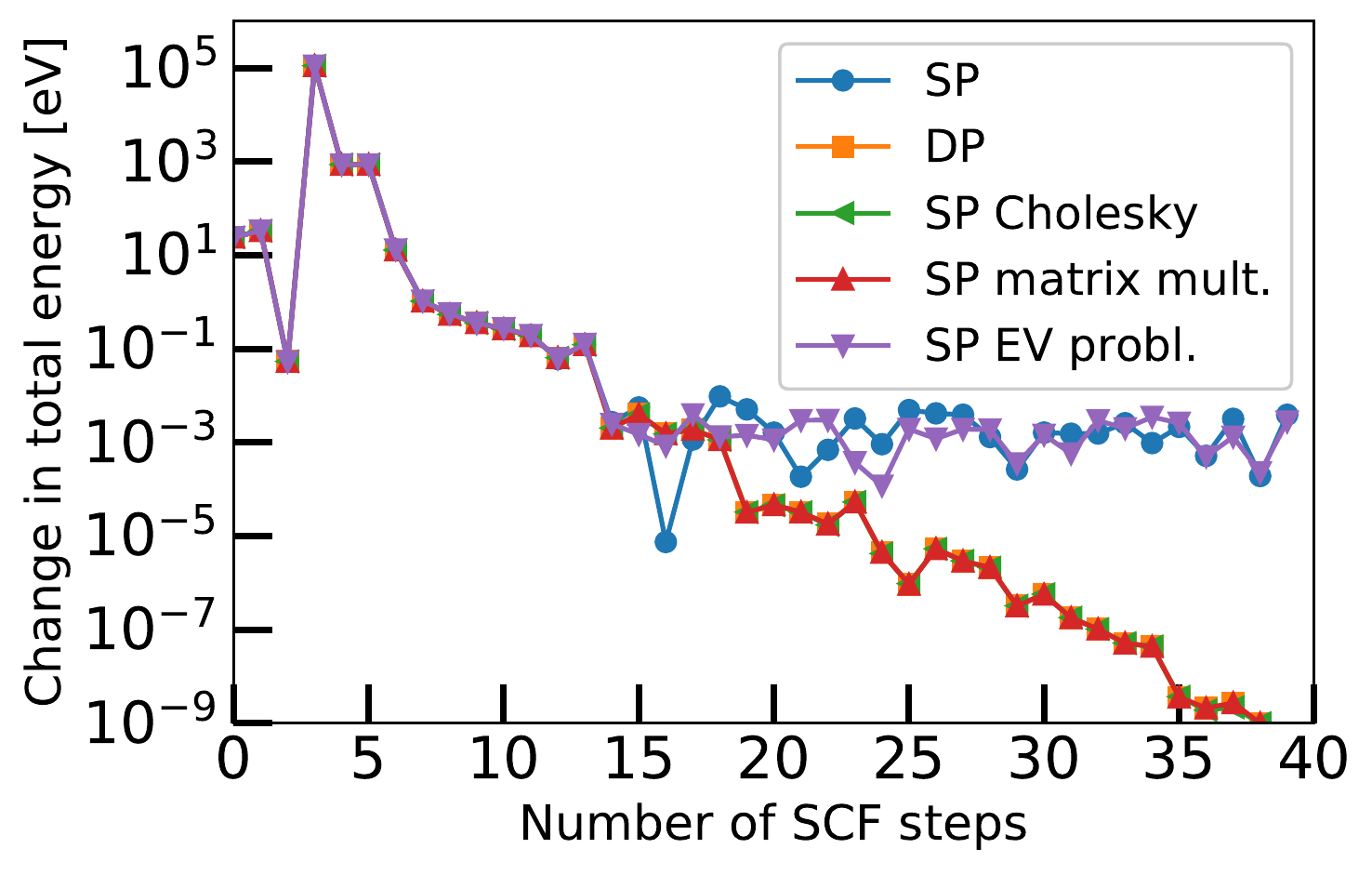}
      & \includegraphics[width=0.45\textwidth]%
                        {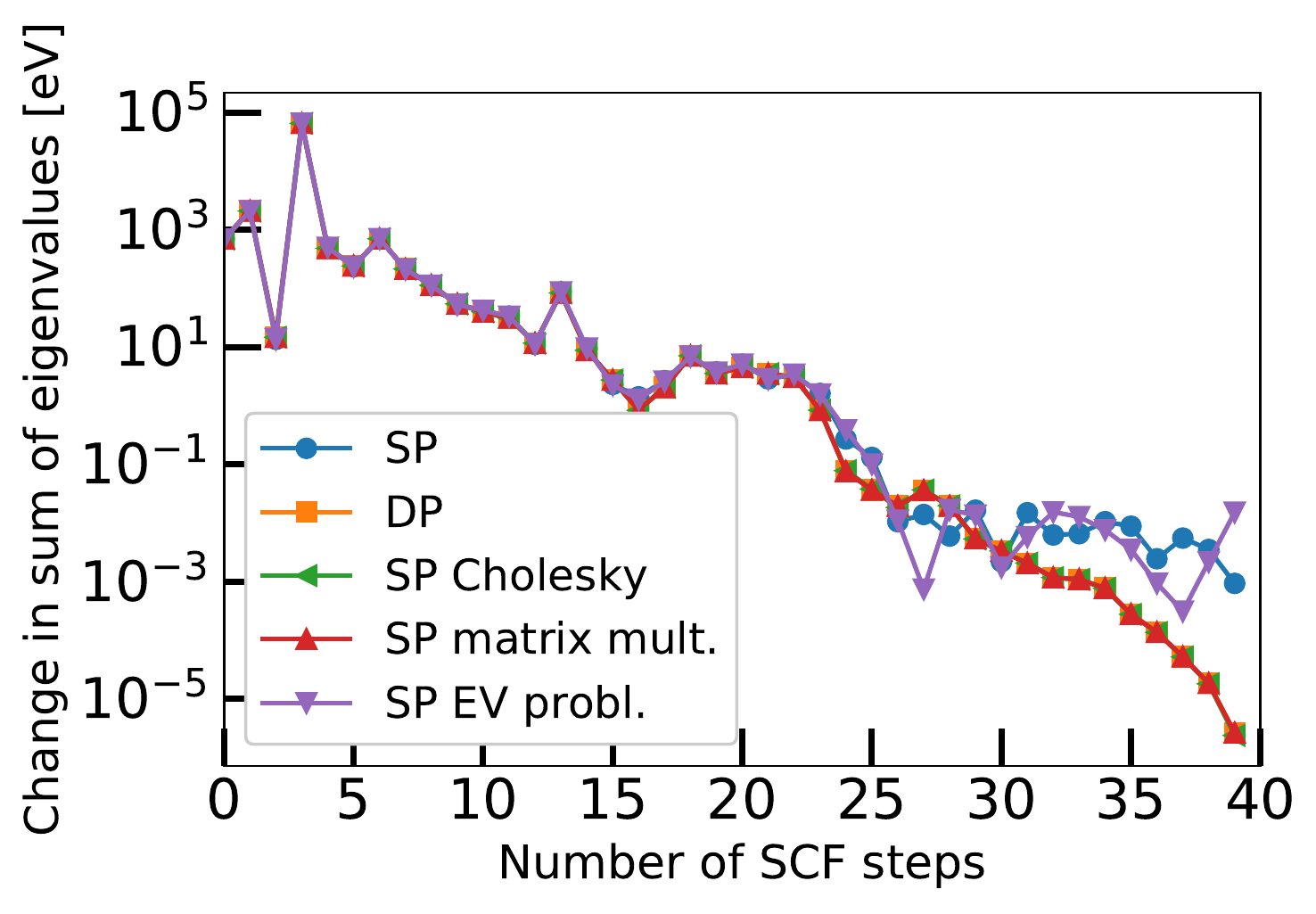}
      \\
        \includegraphics[width=0.45\textwidth]%
                        {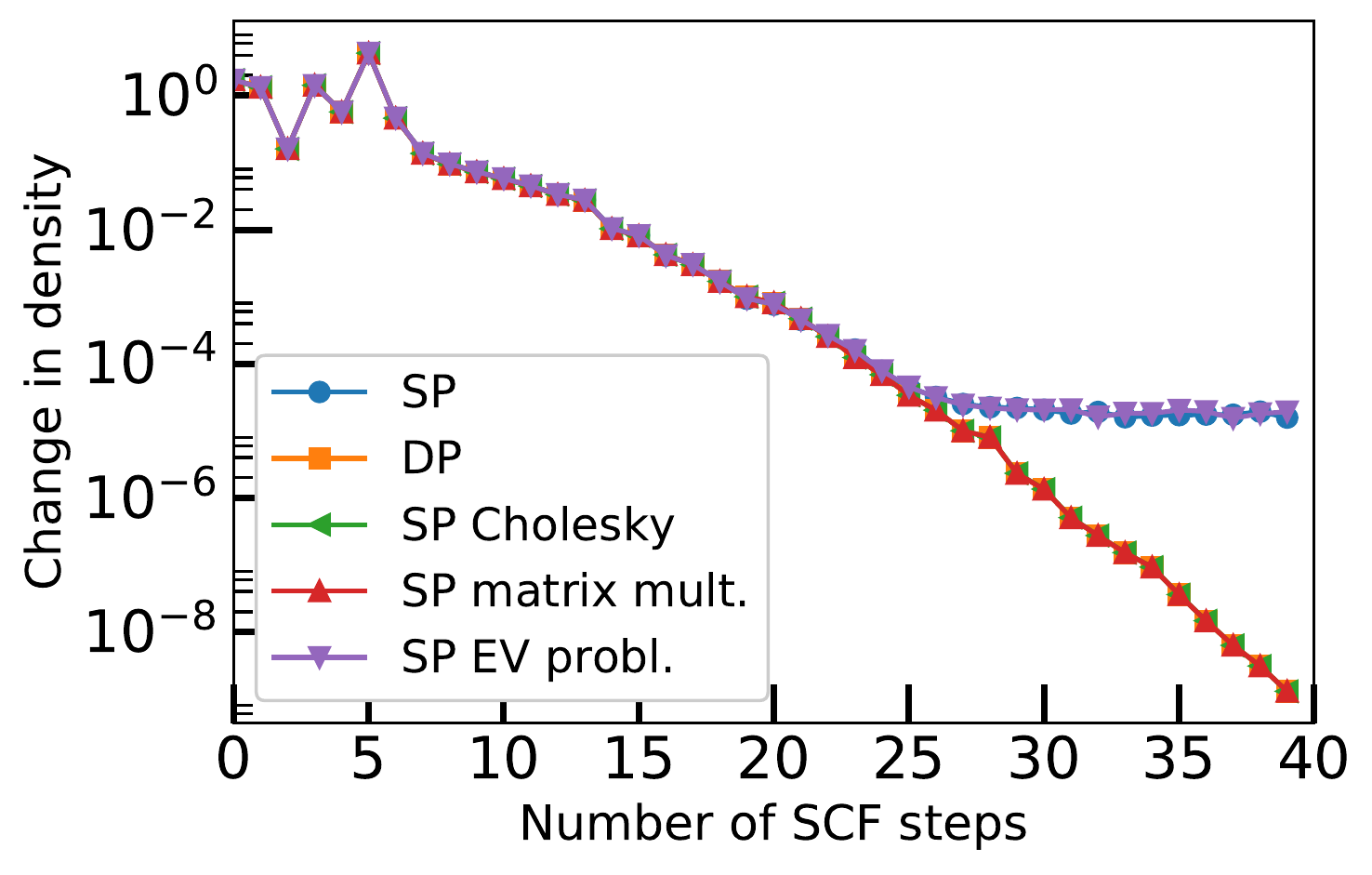}
      & \includegraphics[width=0.45\textwidth]%
                        {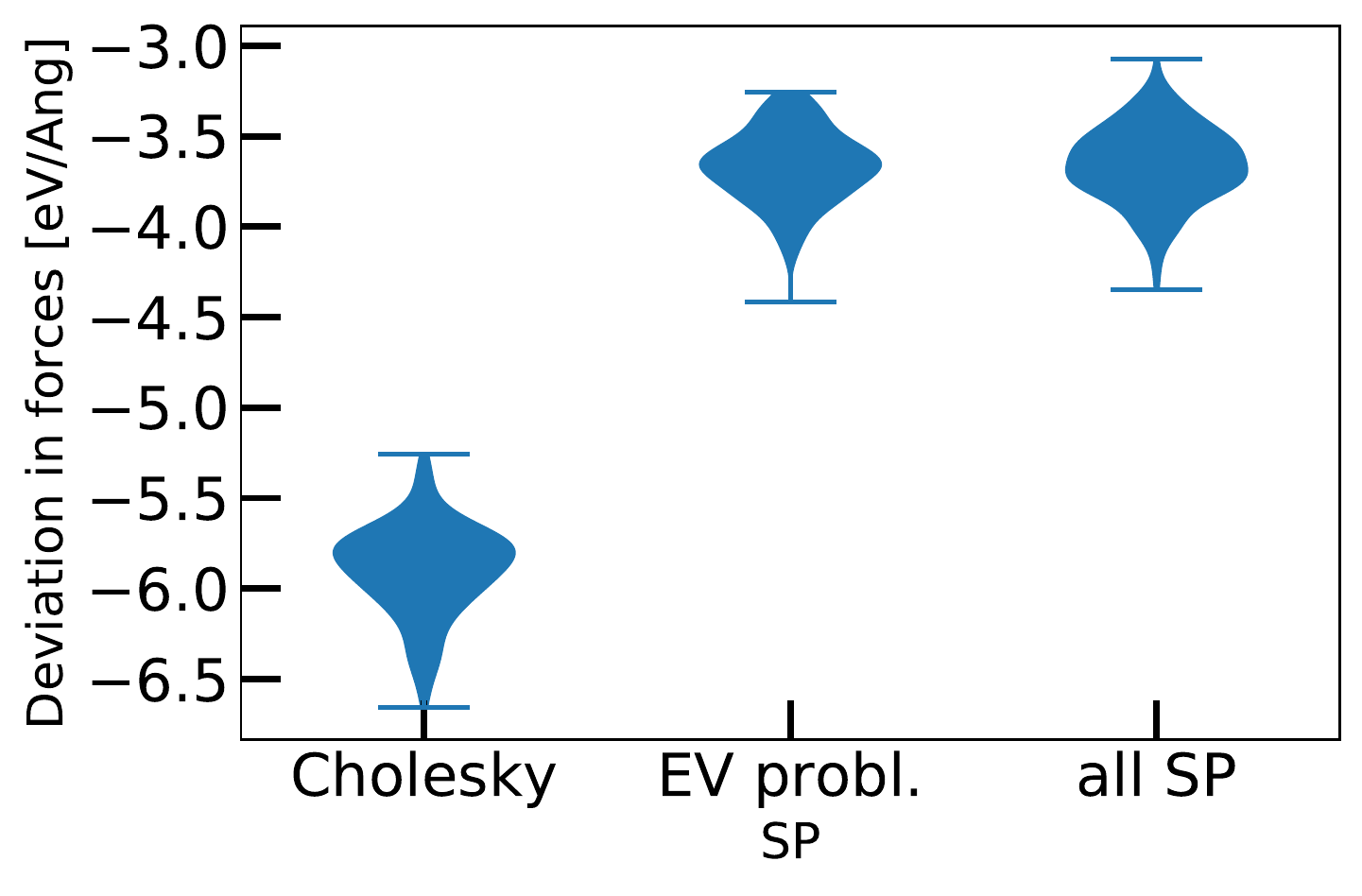}
    \end{tabular}}
  \caption{Convergence behavior of change in total energy (top left),
    change in sum of eigenvalues (top right),
    and change in density (bottom left) with the number of SCF
    iterations for AB-stacked graphite ($M = 108$ atoms,
    $N = 648$ electrons, $K = 23,\!682$).
    SP in the Cholesky decomposition (step (i), green triangles) or the
    matrix multiplication (step (ii), red triangles) show convergence
    identical to full DP (orange squares) calculations and are
    essentially indistinguishable.
    SP in the diagonalization (steps (iii) and (iv), violet
    triangles) corrupts the convergence behavior. 
    A combination of SP in steps (i) through (iv) (blue dots) does
    not reach convergence either. 
    On the bottom right, the deviations in the forces per atom between
    SP and DP calculations are depicted for SP in different steps of the
    solution of the EVP.}
  \label{fig:4}
\end{figure}

In general, not only the central steps~(iii) and (iv) of solving the
EVP (the diagonalization comprising the reduction to tridiagonal form
and the tridiagonal eigensolver; cf.\
  Section~\ref{sec:ELPA-AEO}),
but also the Cholesky decomposition~(i) and the transformation to the
standard eigenproblem~(ii) offer the flexibility to choose between
SP and DP. 
Even though the overlap-matrix $B$ remains constant during an SCF cycle
for an atom-centered basis, the test calculations on an AB-stacked
graphite system ($M = 108$ atoms, PBE exchange-correlation functional,
``tight'' numerical defaults, ``Tier 2'' default basis set,
$K = 23,\!682$ basis functions, $N = 648$ electrons) include the
Cholesky decomposition in every iteration step in order to assess the
impact of SP versus DP in step~(i). 
  Figure~\ref{fig:4}
illustrates that SP in (i) and (ii) does not noticeably change the
convergence behavior of the extensive properties (change of total
energy
$\Delta E [n(\vec{r})] = E [n^{(k)}(\vec{r})]- E [n^{(k+1)}(\vec{r})]$,
eigenvalues $\Delta \varepsilon$, and density $\Delta n$) during one SCF
cycle and hence, full convergence is achieved in contrast to SP in the
diagonalization~(iii) and (iv). 
This is confirmed in the bottom right picture in
   Figure~\ref{fig:4},
where the forces on each atom, i.e.,
the gradients $\vec{F}_I = - \nabla_{\vec{R}_I} E_0(\{\vec{R}\})$ and
their deviation from the full double precision values
$|\vec{F}^{DP}_I - \vec{F}^{SP}_I|$ are shown.
The force per atom, an intensive quantity, is typically monitored and
required to reach a certain accuracy in calculations targeted at
exploring~$E_0(\{\vec{R}\})$.
The bottom right plot in
  Figure~\ref{fig:4}
confirms that SP in the Cholesky decomposition (i) influences the
results only marginally; SP transformation (ii) even yields numerically
identical results (not shown on the logarithmic scale).
By contrast, a SP diagonalization results in force deviations of
up to $0.5$~meV/\AA, which will still be sufficiently small for
certain applications such as prescreening in PES exploration or
statistical methods based on sampling by MD, when interpreting the error
noise in the forces as acceptable thermal noise~%
  \cite{kuhne2007}.
For the combination of SP throughout steps (i) to (iv), the
convergence behavior and the force deviations are dominated by the
performance of the eigensolver steps (iii) and (iv), and the convergence
criteria for neither energy, eigenvalues, nor density are fulfilled. 
However, as discussed for
  Figures~\ref{fig:FHI1} and \ref{fig:FHI2},
resorting to a diagonalization (iii) and (iv) in SP during the initial
SCF steps is computationally advantageous, but switching to DP is
required in the final steps for full convergence.

  Figure~\ref{fig:7}
shows that the discussed advantages of SP are preserved in massively
parallelized computations.
Here, we display calculations for a slab of silicon carbide, where a
layer of graphene is adsorbed on the surface~%
  \cite{Nemec:2013jm}.
Compared to the 2013 ELPA code base, which presents a common usage
scenario before the ELPA-AEO project, we observe a speed-up of $1.7$ for
DP calculations.
Another factor of $1.4$ is obtained when switching to SP, which
would not have been possible with earlier releases of the library.
The  almost ideal strong scaling with respect to the number of cores is
retained in SP calculations.

\section{Concluding remarks}

The ESSEX-II and ELPA-AEO projects are collaborative research efforts
targeted at developing iterative solvers for very large scale
eigenproblems (dimensions $\gg 1\mathrm{M}$) and direct solvers for
smaller-scale eigenproblems (dimensions up to $1\mathrm{M}$), and
at providing software for these methods.
After briefly highlighting some recent progress in the two projects
w.r.t.\ auto-tuning facilities, resilience, and added functionality, we
have discussed several ways of using mixed precision for reducing the
runtime.

In iterative schemes such as BEAST, single precision may be used in
early iterations.
This need not compromise the final accuracy if we switch to double
precision at the right time.
Even working in extended precision may speed up the execution if the
extra precision leads to fewer iterations and is not too expensive,
as seen with an iterative orthogonalization scheme for the block
Jacobi--Davison method.
Additional finer-grained control of the working precision, addressing
just particular steps of the computations can also be beneficial; this
has been demonstrated with electronic structure computations, where the
precision for each step was chosen directly from the calling code.

Our results indicate that the users should be able to adapt the
working precision, as well as algorithmic parameters, to their
particular needs, together with heuristics for automatic selection.
Work towards these goals will be continued in both projects.




\bibliographystyle{spmpsci}      
\bibliography{paper}


\end{document}